\documentclass[10pt,final,journal,letterpaper,twoside,twocolumn]{IEEEtran}
\usepackage{amsmath}                
\usepackage{amssymb}               
\usepackage{amsfonts}               
\usepackage[hidelinks]{hyperref}    
\usepackage{subcaption}
\usepackage{stfloats}
\usepackage{xfrac}
\usepackage{graphicx}               
\usepackage{pgfplots}               
\usepackage{cite}                   
\pgfplotsset{compat=1.15}

\newtheorem{proposition}{Proposition}
\newtheorem{remark}{Remark}

\usepackage{relsize}
\usepackage{mathtools, nccmath}
\newcommand{\yrf}{y_{\mathrm{rf}}}
\newcommand{\yrfi}{y_{I}}
\newcommand{\yrfq}{y_{Q}}
\newcommand{\xrfi}{x_{I}}
\newcommand{\xrfq}{x_{Q}}
\newcommand{\yopt}{y_{O}}
\newcommand{\xopt}{x_{O}}
\newcommand{\dmin}{d_{\mathrm{min}}}
\newcommand{\boldy}{\mathbf{Y}}
\newcommand{\boldx}{\mathbf{X}}
\newcommand{\boldn}{\mathbf{N}}
\DeclarePairedDelimiter{\nint}\lfloor\rceil

\begin{document}

\title{Cross-Band Modulation Design for Hybrid RF-Optical Systems}


\author{Thrassos K. Oikonomou,~\IEEEmembership{Graduate Student Member,~IEEE}, Sotiris A. Tegos,~\IEEEmembership{Senior Member,~IEEE}, \\ Panagiotis D. Diamantoulakis,~\IEEEmembership{Senior Member,~IEEE}, and George K. Karagiannidis,~\IEEEmembership{Fellow,~IEEE}
\thanks{The authors are with the Department of Electrical and Computer Engineering, Aristotle University of Thessaloniki, 54124 Thessaloniki, Greece (e-mails: toikonom@ece.auth.gr, tegosoti@auth.gr, padiaman@auth.gr, geokarag@auth.gr).}
\thanks{This work was supported by the Hellenic Foundation for Research and Innovation (H.F.R.I.) under the “3rd Call for H.F.R.I. Research Projects to support Post-Doctoral Researchers” (Project Number: 7280).}
}
 
\maketitle

\begin{abstract}
This paper presents a novel cross-band modulation framework that integrates three-dimensional (3D) modulation in the RF domain with intensity modulation and direct detection (IM/DD) in the optical domain, marking the first approach to leverage this combination for improved communication reliability. Designed to harness cross-band diversity, the proposed framework optimizes symbol mapping across the RF and optical links, significantly improving mutual information (MI) and symbol error probability (SEP) performance. In this context, we propose two practical cross-band modulation schemes to implement this framework, both of which utilize quadrature amplitude modulation (QAM) in the RF subsystem. The first is a linear cross-band mapping scheme, where RF symbols are mapped to optical intensity values through an analytically tractable optimization, ensuring low complexity with $\mathcal{O}(1)$ detection while minimizing the SEP. The second is a deep neural network generated 3D constellation (DNN-Gen), which uses a custom-designed loss function to optimize symbol placement, maximizing the MI and minimizing the SEP by learning an adaptive symbol mapping function. Although DNN-Gen introduces additional computational complexity compared to the linear mapping approach, it achieves significant performance gains by adapting the 3D constellation to varying signal-to-noise ratio (SNR) conditions. Beyond these practical implementations, we derive a theoretical MI benchmark for the linear mapping scheme, providing important insights into the fundamental limits of RF-optical cross-band communication. Extensive Monte Carlo simulations validate the proposed framework and show that both schemes outperform state-of-the-art cross-band modulation techniques, including cross-band pulse amplitude modulation, with significant performance gains. Moreover, DNN-Gen maintains high performance over a range of RF SNRs, reducing the need for exhaustive training at each operating SNR condition, further enhancing its practical applicability. These results establish the proposed cross-band modulation framework as a scalable, high-performance solution for next-generation hybrid RF-optical networks, offering a balance between low-complexity implementation and optimized symbol mapping to maximize system reliability and efficiency.
\end{abstract}    


\section{Introduction}
Sixth-generation (6G) wireless networks are expected to revolutionize communication systems by improving spectral efficiency, reliability, and sustainability to meet the growing demand for ultra-reliable, high-capacity connectivity \cite{6G_karag,6G}. As data traffic grows, traditional radio frequency (RF) systems that offer wide coverage and adaptability are increasingly constrained by spectrum scarcity, multipath fading, and co-channel interference \cite{rapaport}. At the same time, optical wireless (OW) technologies such as free-space optics (FSO) and visible light communication (VLC) have emerged as promising alternatives, offering high data rates, low latency, and immunity to RF interference \cite{ow-1, ow-2, ow-3}. However, these systems rely on a line-of-sight (LoS) transmission path and are highly susceptible to atmospheric disturbances such as fog, rain, and scintillation, which limit their stand-alone reliability \cite{ow-4, ow-5}.

To mitigate these limitations, hybrid RF-optical communication systems have emerged as a promising approach to leverage the complementary properties of RF and OW channels \cite{hybrid-1, hybrid-2, hybrid-3, Dobre}. By integrating RF and optical transmission, cross-band systems can increase spectral efficiency, improve robustness to channel impairments, and achieve diversity gains by jointly processing signals in both domains \cite{hybrid-4, hybrid-5, hybrid-6, Haas}. Despite these advantages, existing hybrid RF-optical systems predominantly rely on independent signal processing for each band, limiting the potential for joint optimization and efficient cross-band modulation. This highlights the need for a unified modulation framework that can fully harness the benefits of cross-band communication, enabling more efficient signal transmission and improved overall system performance.

\subsection{State-of-the-Art}
Hybrid RF-optical communication systems have been extensively studied to improve reliability and spectral efficiency by leveraging the complementary properties of RF and optical channels. Early approaches explored hard switching, where either the RF or FSO channel is selected for transmission, and soft switching, where both channels are used simultaneously \cite{Hranilovic}. However, hard switching fails to exploit the available diversity gains due to the lack of joint decoding, while soft switching processes RF and optical signals separately before combining them, limiting its ability to fully utilize cross-band diversity. Alternative hybrid RF-optical approaches have integrated RF (sub-6 GHz or mmWave) and optical wireless (OW) transmission through data splitting or codeword partitioning \cite{Nestoras-2, Rallis}. Although these methods enable cross-band communication, data splitting suffers from significant performance degradation when the RF and optical channels have asymmetric SNR conditions, making it inefficient under dynamic channel variations. In addition, \cite{Nestoras-1} analyzed the average bit error rate of a hybrid system with a conventional RF receiver, assuming that phase shift keying (PSK) is used in the optical subsystem. This work also explored selection combining (SC), which uses PSK with a large DC bias to allow transmission via IM/DD in the optical link. However, such an approach is highly inefficient because PSK requires excessive DC bias in the optical domain, which significantly reduces power efficiency. In addition, the authors in \cite{Optik, Hybrid-Alouini, Yue} studied the outage probability for hybrid RF-FSO and RF-VLC systems using SC to achieve diversity gains. Although SC has been extensively investigated in hybrid systems due to its simplicity, it inherently processes RF and optical signals independently, thereby failing to fully realize the diversity potential of cross-band communication.

To overcome these limitations, joint signal processing has been proposed. In \cite{Popovski}, a hybrid receiver architecture was introduced that combines coherent and non-coherent RF reception through magnitude-based mapping to improve reliability. However, this approach remains confined to the RF domain, limiting its effectiveness in optical transmission. Extending this, \cite{sotiris-panos} investigated non-coherent modulations in both RF and optical links and demonstrated diversity gains through joint processing. Despite these advances, the use of non-coherent modulation in both subsystems restricts the ability to fully harness the phase and amplitude characteristics of the transmitted signals. This results in suboptimal symbol detection, reduced spectral efficiency, and an inability to fully exploit the available diversity gains. Furthermore, the separate processing of RF and optical signals prevents a more efficient and adaptive modulation strategy that jointly optimizes symbol mapping across both domains. These drawbacks highlight the need for an alternative approach that fully integrates coherent RF reception with optical IM/DD in a unified cross-band modulation framework.

\subsection{Motivation \& Contribution}
To address these challenges, this work proposes a novel cross-band modulation framework that jointly processes RF and optical signals, leveraging their complementary characteristics to maximize system performance. Unlike existing hybrid RF-optical systems that process signals separately in each domain, our approach unifies coherent RF reception with optical IM/DD to form a three-dimensional (3D) constellation that inherently exploits cross-band diversity. This joint processing strategy not only improves spectral efficiency, but also increases robustness to fading, interference, and dynamic channel variations. In particular, while RF communication provides reliable coverage and resilience to optical impairments, optical IM/DD offers high data rates and immunity to RF interference. However, without a unified modulation strategy that optimally maps symbols across both domains, these advantages remain largely untapped. By integrating coherent RF processing with optical IM/DD in a structured manner, we establish a modulation scheme that efficiently distributes information across the two bands, resulting in improved information transfer, greater diversity gains, and improved adaptability to varying channel conditions.
Thus, to the best of the authors' knowledge, no existing work has proposed a cross-band modulation scheme that integrates coherent RF modulation with IM/DD in the optical domain while jointly optimizing the symbol mapping between the two subsystems to maximize system performance.

To realize this framework, we introduce two practical cross-band modulation schemes: (i) a linear mapping-based cross-band modulation, which provides a tractable and analytically optimized solution with low-complexity detection, and (ii) a deep neural network generated (DNN-Gen) cross-band scheme, which employs deep learning to optimize symbol placement and maximize mutual information (MI) and symbol error probability (SEP) performance. Specifically, the main contributions of this work are as follows:
\begin{itemize}
    \item We propose a unified cross-band modulation framework that integrates coherent RF modulation with IM/DD in the optical domain, marking the first attempt to jointly optimize symbol mapping in both subsystems to leverage cross-band diversity and improve the efficiency of information transmission. This joint processing naturally forms a 3D constellation structure that utilizes the full potential of cross-band transmission.
    \item We introduce a linear cross-band modulation scheme and derive a closed-form MI expression and an analytical SEP expression, offering theoretical insights into the impact of linear mapping on system performance. In addition, we formulate a tractable optimization problem to determine the optimal linear mapping coefficients, ensuring robust and low-complexity cross-band transmission.
    \item We develop a DNN-Gen cross-band modulation scheme that optimizes the 3D constellation structure using a quadrature amplitude modulation (QAM) RF input, generating symbol mappings that maximize the MI and minimize the SEP. To achieve this, we design a custom loss function that increases the packing density of the 3D constellation while enforcing power constraints to ensure optimal performance under different SNR conditions.
    \item We validate the proposed schemes through extensive numerical evaluations, demonstrating that both the linear mapping and DNN-Gen schemes significantly enhance the proposed cross-band framework compared to existing cross-band modulation techniques such as CB-PAM. The linear mapping scheme offers substantial performance gains with an analytically optimized structure and ultra-low complexity detection, making it highly suitable for practical deployment in resource-constrained systems. On the other hand, DNN-Gen further improves MI and SEP performance, achieving more than 2 dB additional gains over the linear scheme by optimizing the symbol placement in both RF and optical domains. However, this comes at the cost of increased computational complexity due to the DNN-based optimization and greater sensitivity to SNR variations, which requires careful adaptation to system conditions.
\end{itemize}

\subsection{Structure}
The remainder of the paper is organized as follows. Section II presents the system model and describes the proposed cross-band modulation framework, detailing the RF and optical signal transmission and the adopted mapping strategies. Section III analyzes the MI and SEP performance of the optimized linear mapping-based cross-band modulation, deriving theoretical expressions under different input distributions to provide analytical insights into the system reliability. Section IV presents the DNN-generated 3D constellation for optimized cross-band modulation, detailing its architecture, custom loss function, and optimization formulation. Section V validates the proposed schemes through Monte Carlo simulations, comparing them to existing cross-band techniques. Section VI concludes with key findings and possible future research directions.

\section{System Model}
This section introduces the proposed hybrid RF-optical communication system, which consists of a transmitter with an RF antenna and an optical source, e.g., light-emitting diode (LED), and a receiver with an RF antenna and a photodiode. The RF link uses two-dimensional (2D) coherent detection, while the optical link uses IM/DD. Both the RF and optical subsystems transmit the same information, with the optical channel data derived directly from the RF channel, thus providing diversity by leveraging their distinct propagation characteristics. This diversity increases reliability and robustness to channel impairments. The joint processing of RF and optical signals at the receiver enables efficient decoding at the combined baseband receiver, maximizing data recovery accuracy and system performance.

\subsection{Signal Model}
The output signals of the RF and optical subsystems are expressed respectively as 
\begin{equation}
\begin{aligned}
    \yrfi = h_1 p_1 \xrfi + n_1 \\
    \yrfq = h_1 p_1 \xrfq + n_2
\end{aligned}
\end{equation}
and
\begin{equation}
\begin{aligned}
    \yopt = h_2 p_2F\left(\xrfi, \xrfq\right) + n_3,
\end{aligned}
\end{equation}
where $\xrfi, \xrfq \in \mathbb{R}$ represent the in-phase and quadrature (I-Q)  components of the 2D coherent modulation used in the RF link, which are assumed to be mutually independent and satisfy $\mathbb{E}\left(\xrfi^2\right) = \mathbb{E}\left(\xrfq^2\right) = \frac{1}{2}$. The function $F(\cdot)$ characterizes the transformation of the I-Q domain symbols into the optical domain, mapping the modulated RF signal to its corresponding optical representation in a way that ensures unit average energy, expressed as $\mathbb{E}\left[F^2\left(\xrfi, \xrfq\right)\right] = 1$. Since the optical subsystem operates under IM/DD, which inherently relies on modulating the intensity of the optical carrier rather than its phase or frequency, the transmitted optical signal must remain strictly non-negative at all times, leading to the fundamental requirement that $F\left(\xrfi, \xrfq\right) \geq 0$ for all possible values of $x_I$ and $x_Q$, thereby ensuring compatibility with the non-coherent nature of optical reception while preserving the integrity of signal detection. Furthermore, in this setup, $p_1$ and $p_2$ denote the transmit powers allocated to the RF and optical links, respectively, and also $h_1$ and $h_2$ represent the power gains associated with the RF and optical channels, respectively. The noise components, modeled as additive white Gaussian noise (AWGN), are introduced into the RF and optical links. The AWGN terms in the RF link, $n_1$ and $n_2$, are modeled as zero-mean Gaussian random variables with variance $\sigma_n^2$, i.e., $n_1, n_2 \sim \mathcal{N}(0, \sigma_n^2)$, and the AWGN in the optical link, $n_3$, is modeled as a zero-mean Gaussian random variable with variance $\sigma_o^2$, i.e., $n_3 \sim \mathcal{N}(0, \sigma_o^2)$.

\subsection{Cross-band Receiver}
In the proposed cross-band system, the receiver jointly processes the 2D RF signal $\yrf = \left(\yrfi, \yrfq\right)$ in the I-Q plane and the one-dimensional optical signal $\yopt$ along the optical power (OP) axis, forming a 3D constellation in I-Q-OP space. This integrated approach contrasts with conventional systems that modulate the RF and optical signals independently using either 2D coherent I-Q modulation or one-dimensional non-coherent optical power modulation. Taking advantage of the correlation between the RF and optical subsystems, the proposed receiver improves the robustness to noise and channel impairments, resulting in more efficient and reliable signal detection.

The receiver operates by defining a unified decision region in 3D space for the $i$-th transmitted symbol, denoted as $\left(x_{I, i}, x_{Q, I}, x_{O, i}\right)$, and both subsystems carry the same information. Therefore, the decision region for the $i$-th symbol, assuming equiprobable symbols and using the maximum likelihood (ML) criterion, is defined as in \cite{sotiris-panos}
\begin{equation} \label{eq:voronoi_det}
\begin{aligned}
\mathcal{V}_{i} \triangleq \left\{(x, y, z)\in \mathbb{R}^3 \!: 
d_i(x, y, z) \leq d_j(x, y, z), \forall j \neq i \right\} \! , 
\end{aligned}
\end{equation}
where the distance metric $d_i(x, y, z)$ is given by
\begin{equation} \label{dmin}
d_i(x, y, z) = \frac{(x - x_{I,i})^2}{\sigma_n^2} + \frac{(y - x_{Q,i})^2}{\sigma_n^2} + \frac{(z - x_{O,i})^2}{\left(1+I_D^2\right)\sigma_o^2}. 
\end{equation}
The decision-making process jointly considers all three dimensions, ensuring accurate decoding through combined baseband processing. It should be noted that the RF and optical symbols may not arrive at the receiver simultaneously due to differences in propagation characteristics between the RF and optical channels, as well as potential synchronization problems. However, this is not a limitation of the proposed system as the received symbols can be buffered and processed together when both components are available. This buffering mechanism ensures that the calculation of the distance metric $d_i$ and the subsequent symbol detection are not affected by timing mismatches.

\section{Linear Cross-Band Modulation}
In this section, we analyze a practical and simple linear mapping for the proposed cross-band modulation scheme, where the optical domain signal is given by
\begin{equation}\label{linear_mapping}
    \begin{aligned}
        F= \frac{1}{\sqrt{1+I_D^2}} \left(a_1\xrfi + a_2\xrfq + I_{D}\right).
    \end{aligned}
\end{equation}
The coefficients $a_1, a_2 \in \mathbb{R}$ are selected to satisfy $a_1^2+a_2^2=2$, ensuring unit average energy when the RF I-Q components are linearly combined to produce the corresponding optical symbol. Since intensity modulation with IM/DD requires a strictly positive optical signal, a DC bias $I_D$ is introduced after the linear combination of the RF components, defined as
\begin{equation}\label{eq:Idc}
\begin{aligned}
    I_{D} = -\min \limits_{\xrfi, \xrfq}\left\{a_1\xrfi + a_2\xrfq\right\},
\end{aligned}
\end{equation}
ensuring that the optical signal remains non-negative.
In the following, we first derive a theoretical MI benchmark for the proposed cross-band system, which provides important insights into the behavior of linear mapping in cross-band modulation. Next, we present a practical implementation for discrete RF inputs with linear mapping, optimizing the mapping coefficients to minimize the SEP and improve system performance. Finally, we derive a closed-form performance analysis for the SEP of this optimized scheme, providing a deeper understanding of its reliability and effectiveness.

\subsection{Mutual Information}
In this subsection, the proposed hybrid system utilizing a linear mapping $F$ is theoretically investigated in terms of MI. The MI between inputs and outputs can be written as
\begin{equation}\label{MI}
\begin{aligned}
    & I(\mathbf{X}; \mathbf{Y}) = h(\mathbf{Y}) - h(\mathbf{Y}|\mathbf{X}) \\
    &\hspace{0.2em}= -\iiint_{\mathbb{R}^3} f_{\mathbf{Y}}(\mathbf{Y})\log_{2}\left(f_{\mathbf{Y}}(\mathbf{Y})\right) d\mathbf{Y} - \frac{1}{2}\log_{2}\left(\mathrm{det}\left(\mathbf{N}\right)\right),
\end{aligned}
\end{equation}
where $\boldx = \left[\xrfi,\xrfq, \xopt\right]^T$, $\boldy = \left[\yrfi, \yrfq, \yopt\right]^T$, $\boldn = \left[n_1,n_2, n_3\right]^T$, $h(\cdot)$ is the entropy of a random variable, and $f_{\mathbf{Y}}(\mathbf{Y})$ is the joint probability density function (PDF) of the outputs $\yrfi$, $\yrfq$, and $\yopt$. After presenting the integral expressions for MI, it is crucial to highlight that obtaining a closed-form solution becomes intractable when the RF input symbols do not follow a Gaussian distribution or when the mapping function $F\left(\xrfi, \xrfq\right)$ is nonlinear. The complexity arises because in such cases the output distribution no longer retains a convenient analytical form, making it impossible to express the MI in a closed, tractable mathematical expression. This fundamental challenge motivates the derivation of a practical closed-form MI expression that, while not necessarily the optimal or tightest upper bound, provides valuable insights into the impact of linear mapping on MI. More specifically, this expression allows us to examine how key system parameters, including the mapping coefficients and channel conditions, influence the achievable MI, providing an analytical benchmark that helps to understand the fundamental trade-offs and performance trends in cross-band linear modulation. 

In this direction, we define the linear Gaussian cross-band (LGCB) modulation, where the RF input symbols follow a Gaussian distribution, which is the optimal input choice for maximizing the MI in the RF link. In LGCB, the RF inputs are Gaussian distributed, and the mapping function $F(x_I, x_Q)$ is linear, ensuring analytical tractability and enabling a closed-form MI expression that provides valuable insights into system performance. Under these conditions, the received signals at both the RF and optical subsystems remain jointly Gaussian, which allows us to express the MI in terms of covariance matrices and compute it efficiently. Based on this, we proceed with the derivation by first establishing the statistical properties of the system, determining the joint PDF of the received signals, and ultimately formulating the MI expression in a closed-form logarithmic representation. In the RF system, the input signals $(x_{I}, x_{Q})$ are independently distributed as Gaussian random variables with zero mean and variance $\sigma_x^2$, i.e., $x_{I}, x_{Q} \sim \mathcal{N}(0, \sigma_x^2)$ \cite{Information-Theory}. Consequently, $y_I$ and $y_Q$ are independent Gaussian random variables characterized by zero mean and variance $h_1^2p_1^2\sigma_x^2 + \sigma_n^2$. Furthermore, $y_O$ is also Gaussian distributed, since $x_O$ is a linear combination of $x_I$ and $x_Q$ with zero mean and variance $\frac{2h_2^2p_2^2}{1+I_D^2}\sigma_x^2 + \sigma_o^2$. Thus, the joint PDF $f_\boldy(\boldy)$ is a multivariate Gaussian with mean $\boldsymbol{\mu}_{Y}$ and covariance matrix $\boldsymbol{K}_Y$ and is expressed as
\begin{equation}\label{jointpdfY}
    \begin{aligned}
        f_\boldy(\boldy) = C\exp{\left(-\frac{1}{2}\left(\boldy-\boldsymbol{\mu}_Y\right)^{T}\mathbf{K}_{Y}^{-1}\left(\boldy-\boldsymbol{\mu}_Y\right)\right)},
    \end{aligned}
\end{equation}
where $C = 1/\left(\left(2\pi\right)^{3/2}\mathrm{det}(\mathbf{K}_{Y})^{1/2}\right)$, $\boldsymbol{\mu}_{Y} = [0,0,I_{D}]$, and
\begin{equation*}
    \begin{aligned}
    \mathbf{K}_Y = \begin{bmatrix}
    c_1^2\sigma_x^2 + \sigma_n^2 & 0 & c_ta_1\sigma_x^2  \\
    0 & c_1^2\sigma_x^2 + \sigma_n^2 & c_ta_2\sigma_x^2 \\
    c_ta_1\sigma_x^2 & c_ta_2\sigma_x^2 & 2c_2^2\sigma_x^2 + \sigma_o^2
    \end{bmatrix},
    \end{aligned}
\end{equation*}
where $c_1 = h_1p_1$, $c_2 = h_2p_2$ and $c_t = \frac{c_1c_2}{\sqrt{1+I_{D}^2}}$.
Given that the joint PDF of the outputs follows a Gaussian distribution, the MI for the proposed system can be expressed as in \cite{Information-Theory}
\begin{equation}\label{mi_of_Gaussian}
    \begin{aligned}
        I(\mathbf{X}; \mathbf{Y}) = \frac{1}{2}\log_{2}\left({\frac{\mathrm{det}\left(\mathbf{K}_Y\right)}{\mathrm{det}\left(\mathbf{N}\right)}}\right),
    \end{aligned}
\end{equation}
and by substituting \eqref{jointpdfY} into \eqref{MI} and performing some algebraic manipulations, the MI can be reformulated as
\begin{equation}\label{mi_linear_mapping_final}
    \begin{aligned}
        &I(\mathbf{X}; \mathbf{Y}) = \frac{1}{2}\log_{2}\left( \frac{\sigma_x^2 + \sigma_n^2 
        }{\sigma_0^2 \sigma_n^4}\right) \\
        &+ \frac{1}{2}\log_{2}\left(\frac{\sigma_x^2\sigma_o^2 + \sigma_n^2\sigma_o^2 + 2c_t^2\sigma_x^2\sigma_n^2}{\sigma_0^2 \sigma_n^4} \right) \!.
    \end{aligned}
\end{equation}

\begin{remark}
If the input signal is continuously Gaussian distributed, \eqref{mi_linear_mapping_final} shows that the MI remains independent of the parameters $a_1$, $a_2$. This can be understood by noting that $x_I$ and $x_Q$ are independent Gaussian random variables, and the mapping coefficients $a_1$ and $a_2$ merely linearly combine these components without changing their fundamental statistical properties. Since the total signal power is preserved under the condition $a_1^2+a_2^2=2$, the MI depends only on the SNRs of the RF and optical subsystems and not on how the Gaussian-distributed inputs are projected onto the optical link. Furthermore, an important observation from \eqref{mi_linear_mapping_final} is that the MI degrades as the DC bias increases. This occurs because the transmitted optical signal must be normalized in a linear optical mapping to maintain a unit average optical power. As a result, greater DC bias effectively reduces the available dynamic range of the optical link, limiting its ability to faithfully transmit variations in the input signal. Consequently, excessive DC bias reduces overall system performance, which must be considered when optimizing cross-band modulation schemes.
\end{remark}

While a Gaussian input maximizes MI in the RF link and serves as a well-suited input distribution for cross-band transmission when the SNR of the RF subsystem is dominant over the optical SNR, it is not necessarily the optimal choice when the optical SNR significantly exceeds the RF SNR. In such cases, MI can be further improved by adjusting the input distribution to better utilize the capacity of the optical channel. As noted in \cite{wigger}, the optimal input distribution for maximizing MI at the optical receiver follows an exponential distribution. Since the optical input is obtained by a linear combination of the RF I-Q components, achieving an exponential optical signal requires that the RF input components follow a chi-squared distribution, leading to what we define as linear exponential cross-band (LXCB) modulation. Specifically, in LXCB, the RF input components $x_I$ and $x_Q$ are modeled as independent $0.577\sigma_x^2\chi^2(1)$ random variables, so that the average energy in each RF dimension is $\sigma_x^2$ and their transformation through the linear mapping function $F(x_I, x_Q)$ results in an exponential distribution at the optical input. This fully utilizes the optical SNR, enabling a higher MI compared to Gaussian-distributed RF inputs when the optical link has better SNR conditions than the RF link. A key advantage of this approach is that it naturally produces a strictly positive optical signal, eliminating the need for DC bias, i.e., $I_D=0$. In contrast, when the RF inputs are Gaussian distributed, their negative tails require a large DC bias to shift the optical signal into a strictly positive range, resulting in a loss of dynamic range and additional normalization constraints. The gamma-distributed RF input in LXCB modulation inherently avoids this problem, ensuring that all optical signal values remain positive after linear mapping, thus maximizing the capacity of the optical link without the need for further adjustments. As a result, LXCB modulation efficiently leverages high optical SNR scenarios, although its theoretical investigation remains challenging, as a closed-form expression for MI cannot be obtained from \eqref{MI} in this case due to the intractability of the resulting integral expressions.

\subsection{A Practical Linear Cross-band Modulation Scheme}

\begin{figure}
\centering
\begin{tikzpicture}[thick, scale=0.9]
\begin{axis}[
    view={60}{35},          
    xlabel={$x$-axis},      
    ylabel={$y$-axis},      
    zlabel={$z$-axis},      
    xtick style={draw=none}, 
    ytick style={draw=none}, 
    ztick style={draw=none}, 
    domain=-1:1,            
]
\addplot3+[
    only marks,
    mark options={black},
    mark=*, 
    mark size=2pt,          
] table {figures/3D_constellations/3D_linear.txt};

\addplot3[
    thick, 
    color=black
] coordinates {
    (-0.948, -0.948, 0) (-0.948, 0.948, 0.884)
};
\addplot3[
    thick, 
    color=black
] coordinates {
    (-0.948, -0.948, 0) (0.948, -0.948, 0.884)
};
\addplot3[
    thick, 
    color=black
] coordinates {
    (0.948, -0.948, 0.884) (0.948, 0.948, 1.769)
};

\addplot3[
    thick, 
    color=black
] coordinates {
    (0.948, 0.948, 1.769) (-0.948, 0.948, 0.884)
};

\addplot3[
    thick, 
    color=black
] coordinates {
    (-0.948, 0.316, 0.589) (0.948, 0.316, 1.474)
};

\addplot3[
    thick, 
    color=black
] coordinates {
    (-0.948, -0.316, 0.294) (0.948, -0.316, 1.179)
};

\addplot3[
    thick, 
    color=black
] coordinates {
    (0.316, -0.948, 0.589) (0.316, 0.948, 1.474)
};

\addplot3[
    thick, 
    color=black
] coordinates {
    (-0.316, -0.948, 0.294) (-0.316, 0.948, 1.179)
};

\addplot3[
    only marks,
    mark=*,
    mark options={black},
    mark size=2pt
] coordinates {
    (0.3,-0.8,2) 
};
\addplot3[
    thick, 
    color=black
] coordinates {
    (0.3,-0.8,2) (-0.5, -0.25, 0)
};
\node at (0.3,-0.8,2.1) [anchor=south] {$\mathbf{r}$};
\addplot3[
    only marks,
    mark=star,
    mark options={black},
    mark size=2pt
] coordinates {
    (-0.5, -0.25, 0) 
};
\end{axis}
\end{tikzpicture}
\caption{Proposed 3D cross-band constellation with linear mapping.}
\label{fig:3d_plot}
\end{figure}
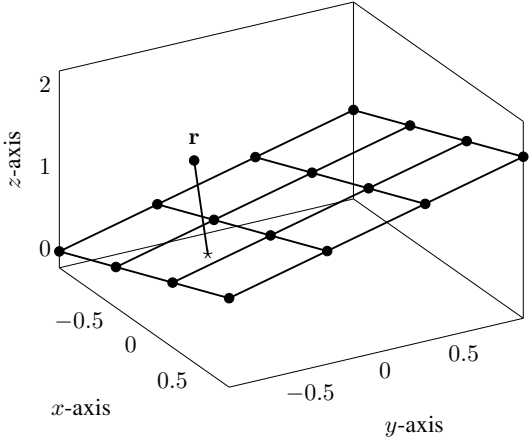

In the proposed practical cross-band scheme, the RF subsystem transmits an $M$-QAM constellation with I-Q coordinates $(x_I,x_Q)$, while the optical symbol is generated as a linear combination of these coordinates. As a result, the symbols of the joint 3D constellation lie on a 2D plane in 3D space. Specifically, the coordinates of a 3D symbol are given by $\left(x_I,x_Q, \frac{a_1x_I+a_2x_Q + I_D}{\sqrt{1+I_D^2}}\right)$, where the plane on which these symbols lie is described by the equation $z = a_1x_I + a_2x_Q$. 

Since all constellation symbols are confined to a single plane in 3D space, the detection process for the proposed cross-band modulation scheme becomes significantly more efficient. Instead of iteratively computing the detection metric defined in \eqref{dmin} and performing $M$ distance comparisons between the received symbol $(y_I, y_Q, y_O)$ and the symbols of the 3D constellation, the detection can be streamlined by projecting the received 3D symbol onto the corresponding 2D plane, as shown in Fig. \ref{fig:3d_plot}. By utilizing the parallelogram structure inherent in the constellation on the 2D plane, the detection complexity is reduced to $\mathcal{O}(1)$ by using predefined thresholds corresponding to the 2D grid, thus greatly simplifying the computational requirements. 

To elaborate, in standard $M$-QAM, the constellation points are arranged in a rectangular lattice structure that lies entirely on the I-Q plane. If we denote the planes along the I-axis as $\left\{x_I = k_I \Delta_I | k_I\in \mathbb{Z}\right\}$ and along the Q-axis as $\left\{x_Q = k_Q \Delta_Q | k_Q\in \mathbb{Z}\right\}$, then each constellation symbol is positioned at coordinates $(x_I, x_Q) = (k_I\Delta_I, k_Q\Delta_Q)$ forming a uniform grid with constant spacing of $\Delta_I > 0$ on the I-axis and $\Delta_Q > 0$ on the Q-axis, where $k_I$ and $k_Q$ are the integer indices of the constellation points along the I-Q axis. Although the received symbol $(y_I, y_Q, y_O)$ is in 3D space, the fact that all constellation points lie on a well-defined 2D plane allows us to project the received symbol onto this plane. The resulting projected point is expressed as $P=(X, Y, a_1X+a_2Y)$, where the coordinates $X$ and $Y$ correspond to the projections of the received symbol on the axes of the 2D plane. The squared distance between this projected point and any point on the 2D grid is then calculated as
\begin{equation} \label{d2}
    \begin{aligned}
        d_{P}^2 = &\frac{\left(X-k_I \Delta_I\right)^2}{\sigma_n^2} + \frac{\left(Y-k_Q \Delta_Q\right)^2}{\sigma_n^2}\\
        &+ \frac{\left(a_1\left(X-k_I\Delta_I\right) + a_2\left(Y-k_Q\Delta_Q\right)\right)^2}{\left(1+I_D^2\right)\sigma_o^2}.
    \end{aligned}
\end{equation} 
\begin{proposition}
    The optimal indices $\left(\hat{k}_I,\hat{k}_Q\right)$ that minimize the squared distance $d_{k_I,k_Q}^2$ between a constellation point and the received symbol $P$ are determined by
    \begin{equation} \label{ind}
        \begin{aligned}
            \hat{k}_I = \nint[\Big]{\frac{X}{\Delta_I}},\hspace{0.5em}
            \hat{k}_Q = \nint[\Big]{\frac{Y}{\Delta_Q}},
        \end{aligned}
    \end{equation}
    where $\nint{\cdot}$ denotes the rounding to the nearest integer.
\end{proposition}
\begin{IEEEproof}
    To determine the minimizing pair $\left(\hat{k}_I,\hat{k}_Q\right)$,
    the integer variables $(k_I,k_Q)$ are treated as continuous variables. The solution to this relaxed continuous problem is then rounded to the nearest integer to obtain the optimal solution for the integer case. Since the squared distance in \eqref{d2} is a quadratic function of $k_1$ and $k_2$, its minimum can be found by setting the partial derivatives of $d_{P}^2$ with respect to $k_I$ and $k_Q$ to zero. By computing these derivatives, we derive the following system of equations
    \begin{equation}\label{F}
        \begin{aligned}
            \frac{X-k_I\Delta_I}{\sigma_n^2} & = -\frac{a_1S}{\sigma_o^2} \\
            \frac{Y-k_Q\Delta_Q}{\sigma_n^2} & = -\frac{a_2S}{\left(1+I_D^2\right)\sigma_o^2},
        \end{aligned}
    \end{equation}
    where $S = a_1\left(X-k_I\Delta_I\right) + a_2\left(Y-k_Q\Delta_Q\right)$. From \eqref{F}, it is evident that both equations are satisfied when $S=0$. Substituting this into the equations, the solution simplifies to \eqref{ind},
    which completes the proof. 
\end{IEEEproof}
\begin{remark}
    Although the plane is tilted by parameters $a_1$ and $a_2$, the indices of the nearest lattice points remain unaffected, relying solely on simple divisions and rounding. Consequently, the detection complexity remains $\mathcal{O}(1)$, just like standard QAM in an RF system. However, by leveraging the diversity of two separate links, this cross-band configuration places the lattice points in a 3D plane, adding another non-negative component to the minimum distance. As a result, the minimum distance increases, improving the SEP, without increasing the detection complexity.
\end{remark}

Furthermore, the squared distance between two points $P_i$, $P_j$ in the 3D constellation lattice is given by 
\begin{equation}\label{d_k1,k2}
    \begin{aligned}
        d_{k_1,k_2}^2 = \frac{k_1^2\Delta_I^2}{\sigma_n^2} + \frac{k_2^2\Delta_Q^2}{\sigma_n^2} + \frac{(a_1 k_1\Delta_I - a_2 k_2 \Delta_Q)^2}{\left(1+I_{D}^2\right)\sigma_o^2},
    \end{aligned}
\end{equation}
where $k_1 = k_{I, i} - k_{I,j}$ is the distance between the two points along the I-axis and $k_2 = k_{Q, i} - k_{Q,j}$ represents the distance along the Q-axis. It is important to note that $k_1$ and $k_2$ cannot be equal to zero at the same time, as this would correspond to the same point being compared to itself. It can be observed from \eqref{d_k1,k2} that under certain noise conditions in the proposed cross-band system characterized by $\sigma_n^2$ and $\sigma_o^2$, the value of $d_{k_I,k_Q}^2$ varies depending on the parameters $\left(a_1, a_2, k_1, k_2\right)$. To this end, for a given noise scenario, there exists an optimal set of parameters, denoted $\left(a_1^*, a_2^*, k_1^*, k_2^*\right)$, that maximizes the minimum distance within the 3D constellation lattice and thus minimizes the SEP. To determine the optimal set, we first note that $a_1k_1$ and $a_2k_2$ must have opposite signs to minimize the third term in \eqref{d_k1,k2}. By leveraging symmetry, we restrict our analysis to $a_1, a_2, k_1\geq 0$ and $k_2\leq0$, and under this constraint, the formulated optimization problem that maximizes the minimum distance of the 3D constellation can be described as
\begin{equation*}\tag{\textbf{P1}}\label{eq:maxmin_with_theta}
\begin{aligned}
    \begin{array}{cl}
    \mathop{\mathrm{max}}\limits_{\theta}
    & \mathop{\mathrm{min}}\limits_{k_1, k_2} \mathlarger{d_{k_1,k_2}^2(\theta)} \\
    
    \text{\textbf{s.t.}}& \mathrm{C}_1: \theta\in\left[0,\frac{\pi}{2}\right], \\
    & \mathrm{C}_2:  k_1, k_2 \in \{0, 1, \dots, \sqrt{M} - 1\}, \\
    & \mathrm{C}_3:  k_1 + k_2 \geq 1,
    \end{array}
\end{aligned}
\end{equation*}
where 
\begin{equation}
    \begin{aligned}
        d_{k_1,k_2}^2(\theta) &= \frac{k_1^2\Delta_I^2}{\sigma_n^2} + \frac{k_2^2\Delta_Q^2}{\sigma_n^2} \\
        &+ \frac{(\sqrt{2}k_1\Delta_I\cos{\theta} - \sqrt{2} k_2 \Delta_Q\sin{\theta})^2}{\left(1+I_{D}^2\right)\sigma_o^2}.
    \end{aligned}
\end{equation}
It should be noted that in \eqref{eq:maxmin_with_theta} we have substituted $a_1 = \sqrt{2}\cos{\theta}$ and $a_2 = \sqrt{2}\sin{\theta}$ since $a_1^2 + a_2^2 = 2$ and constraint $\mathrm{C}_3$ is included to ensure that $k_1$ and $k_2$ are never both zero at the same time. 

From \eqref{eq:maxmin_with_theta} it is clear that when the SNR of the RF subsystem significantly exceeds the optical SNR, the minimum distance is primarily determined by the term $\frac {k_1^2}{\sigma_n^2} + \frac{k_2^2}{\sigma_n^2}$. Under these conditions, the optimal pairs $(k_1^*,k_2^*)$ that minimize the distance are $(1,0)$ and $(0,1)$. To maximize the minimum distance, the expression $\left(\sqrt{2}k_1\cos{\theta} - \sqrt{2}k_2\sin{\theta}\right)^2$ must be maximized for both pairs, which is achieved when $\theta^* = \frac{\pi}{4}$. Conversely, when the optical SNR dominates the RF SNR, the term $\left(\sqrt{2}k_1\cos{\theta} - \sqrt{2}k_2\sin{\theta}\right)^2$ becomes the primary determinant of the minimum distance. In this scenario, the minimum distance is minimized when $k_1\cos{\theta} - k_2\sin{\theta}\rightarrow 0$. To increase the minimum distance, it is advantageous that both $k_1$ and $k_2$ are non-zero, since this increases the value of $\frac {k_1^2}{\sigma_n^2} + \frac {k_2^2}{\sigma_n^2}$, thus improving the overall minimum distance. Consequently, the optimal angle $\theta^*$ is the one that makes $k_1\cos{\theta} - k_2\sin{\theta}\rightarrow 0$, ensuring the maximum possible minimum distance between constellation points. In conclusion, the derivation of the optimal triple $\left(\theta^*,k_1^*,k_2^*\right)$ can be efficiently achieved by a straightforward approach that involves partitioning the interval $\left[0,\frac{\pi}{2}\right]$ into small segments and performing a brute-force search to solve \eqref{eq:maxmin_with_theta}. This method ensures a reliable determination of the optimal parameters while maintaining simplicity in implementation.

\subsection{Performance Analysis}
In this subsection, we derive a simple closed-form approximation for the SEP of the proposed modulation scheme. The proposed constellation forms a parallelogram-shaped QAM lattice on a 2D plane within the 3D space, whose structure varies according to the SNR conditions. This variation occurs because the optimal angle $\theta^*$ depends on the specific SNR conditions in each subsystem. In addition, the noise on this 2D plane has unit variance because the detection metric used to make decisions is a weighted Euclidean distance that normalizes the variances of the AWGN components across all dimensions of the I-Q-OP space to unity.

\begin{proposition}
    The SEP of the proposed cross-band modulation scheme can be approximated by
\begin{equation}\label{SEP_approx}
    \begin{aligned}
        P_s & \approx 1 - \left(1 - A_1Q\left(\sqrt{\frac{3A_2\gamma_1^2 + 6\gamma_2^2\frac{\left(A_2-A_3\right)}{1+I_D^2}}{2\left(M-1\right)}}\right)\right)\\
        & \quad \times \left(1 - A_1Q\left(\sqrt{\frac{3A_2'\gamma_1^2 + 6\gamma_2^2\frac{\left(A_2'-A_3'\right)}{1+I_D^2}}{2\left(M-1\right)}}\right)\right),
    \end{aligned}
\end{equation}
where $Q(\cdot)$ denotes the Q-function, $A_1 = 2\left(1-1/\sqrt{M}\right)$, $A_2 = {k_1^*}^2 + {k_2^*}^2$,  $A_3 = 2k_1^*k_2^*\cos{\theta^*}\sin{\theta^*}$, while $\gamma_1^2 = h_1^2p_1^2/\sigma_n^2$ and $\gamma_2^2 = h_2^2p_2^2/\sigma_o^2$ are the received SNRs of each subsystem. In addition, parameters $A_2' = {k_1'^*}^2 + {k_2'^*}^2$ and
$A_3' = 2k_1'^*k_2'^*\cos{\theta^*}\sin{\theta^*}$, where $k_1^*$ and $k_2'^*$ correspond to the pair $(k_1,k_2)$ that yields the second minimum distance of the 3D lattice for a given $\theta^*$. It is worth emphasizing that if the 3D lattice forms a square structure, then the two minimum distances are equal, so that $d_{k_1^*,k_2^*}(\theta^*)=d_{k_1'^*,k_2'^*}(\theta^*)$.
\end{proposition}

\begin{IEEEproof}
    The proposed constellation defines a parallelogram lattice on a 2D plane in 3D space. This allows it to be treated as an $M$-QAM constellation on a 2D plane, characterized by a minimum distance $d_{k_1^*,k_2^*}(\theta^*)$ given in \eqref{d_k1,k2} along one dimension of the 2D plane and distance $d_{k_1'^*,k_2'^*}(\theta^*)$ along the other dimension of the 2D plane. The optimal parameters $(\theta^*,k_1^*,k_2^*)$ that maximize the minimum distance $d_{k_1^*,k_2^*}(\theta^*)$ can be obtained by solving \eqref{eq:maxmin_with_theta}, taking into account the specific SNR conditions. To approximate the SEP, we first consider that the SEP for a 2D $M$-QAM constellation is dominated by the pairwise distance between symbols. Thus, the SEP can be expressed as \cite{proakis}
    \begin{equation}\label{SEP_dmin}
        \begin{aligned}
            P_s \approx 1 &- \left(1 - A_1Q\left(\frac{d_{k_1^*,k_2^*}(\theta^*)}{2}\right)\right)\\
            &\times\left(1 - A_1Q\left(\frac{d_{k_1'^*,k_2'^*}(\theta^*)}{2}\right)\right).
        \end{aligned}
    \end{equation}
    Next, by substituting the optimal parameters in \eqref{d_k1,k2}, we derive the corresponding minimum distance, and by performing some simple algebraic manipulations, we arrive at the approximated SEP expression given in \eqref{SEP_approx}, which completes the proof.
\end{IEEEproof}

\begin{remark}
    It is clear that the SNR conditions affect the argument of the Q-function, as the optimal parameters $(\theta^*, k_1^*, k_2^*)$ that determine the minimum distance vary accordingly. However, as already mentioned, in scenarios where the SNR of the RF subsystem is comparable or higher than that of the optical subsystem, the minimum distance is achieved with a fixed set of parameters, specifically $(\theta^*, k_1^*, k_2^*) = \left(\frac{\pi}{4}, 1, 0\right)$. Under these conditions, the coefficients $A_2$, $A_3$ remain independent of the SNR values in each subsystem, and thus the SEP can be upper-bounded as
    \begin{equation}
        \begin{aligned}
            P_s \leq 1 - \left(1 - A_1Q\left(\sqrt{\frac{3\gamma_1^2 + \frac{6\gamma_2^2}{1+I_D^2}}{2\left(M-1\right)}}\right)\right)^2.
        \end{aligned}
    \end{equation}
\end{remark}
\begin{remark}
    It can be observed that when $A_2 = A_3$, the SEP of the proposed 3D cross-band constellation becomes independent of the optical SNR. Specifically, in the high optical SNR scenario, if we set $\theta=\frac{\pi}{4}$, for the mapping between the RF and optical links, it follows from \eqref{d_k1,k2} that the minimum distance is achieved with the pairs $(k_1, k_2) = (1,0)$ or $(k_1,k_2) = (0,1)$, resulting in $A_2 = A_3$. Under these conditions, regardless of how high the optical SNR becomes, the system performance is entirely determined by the received RF SNR.
    This phenomenon occurs because, for $\theta=\frac{\pi}{4}$, certain RF symbols in the $M$-QAM constellation are mapped to identical optical intensity values. Consequently, even at arbitrarily high optical SNR, the mapping redundancy limits the system performance to that dictated by the RF lattice structure. This finding underscores the critical importance of leveraging the diversity provided by the optical link. To take full advantage of high optical SNR, it is essential to ensure that each RF symbol is uniquely mapped to a distinct optical symbol, thus eliminating redundancy and maximizing the potential performance of the cross-band system.
\end{remark}

\section{DNN-Generated Cross-Band Modulation}
In this section, we optimize the 3D constellation lattice for the proposed cross-band modulation scheme to improve both MI and SEP. To achieve this, we introduce a novel DNN-based approach that takes an $M$-QAM constellation as input to ensure compatibility with existing RF systems. The optimization is performed by learning an improved representation of the constellation through the mapping function $F(x_I, x_Q)$, which determines the transformation of the RF symbols into the optical domain. Through an appropriately designed loss function, the DNN effectively generates 3D constellations that satisfy the energy constraint while optimizing the geometric structure of the constellation to improve the performance of the cross-band system in practical deployment scenarios.

To establish the basis for the optimization, we first analyze the MI of the proposed cross-band system. Since both the input and output signals are discrete, the MI is given by \cite{Information-Theory}
\begin{equation} \label{eq::I_DD}
    \begin{aligned}
        I_{\mathrm{DD}}\left(C;\hat{C}\right) &= H(\hat{C}) - H(\hat{C}|C) \\
        &=\log_2M + \sum_{j=1}^{M}\sum_{i=1}^{M} P\left(\hat{c}^{(j)}|c^{(i)}\right)P\left(c^{(i)}\right)\\
        &\quad \times \log_2\left(\frac{P\left(\hat{c}^{(j)}|c^{(i)}\right)}{\sum_{i=1}^{M}P\left(\hat{c}^{(j)}|c^{(i)}\right)} \right),
    \end{aligned}
\end{equation}
where $C$ and $\hat{C}$ are symbols of the 3D constellation lattice, $\hat{C}$ denotes the decision made by \eqref{eq:voronoi_det}, $P\left(c^{(i)}\right)$ is the probability of transmitting the $i$-th symbol of the 3D lattice, and $P\left(\hat{c}^{(j)}|c^{(i)}\right)$ is the pairwise error probability (PEP) that symbol $\hat{c}^{(j)}$ of the 3D constellation is detected when the symbol $\hat{c}^{(i)}$ is transmitted. Since minimizing the PEP directly improves the reliability of information transmission, and since the PEP is fundamentally related to the geometric arrangement of the 3D constellation, it follows that maximizing the MI is closely related to minimizing the PEP. This connection allows us to establish a natural transition from the MI optimization to the optimization of the geometric properties of the 3D constellation lattice. In particular, since the PEP is largely determined by the minimum Euclidean distance between constellation points, minimizing the PEP can be effectively achieved by maximizing this minimum distance. Thus, the MI optimization of the discrete-input discrete-output cross-band system can be equivalently expressed as maximizing the minimum Euclidean distance in the 3D lattice, making it a core design criterion for improving the MI of the system.

Following the MI analysis, we present a general expression for the SEP, which provides a direct measure of the reliability of the system, and it is written as
\begin{equation} \label{eq:SEP_GENERAL}
    \begin{aligned}
        \mathrm{SEP} = \sum_{j=1}^{M}\sum_{i=1}^{M} P\left(\hat{c}^{(j)}|c^{(i)}\right)P\left(c^{(i)}\right).
    \end{aligned}
\end{equation}
As can be observed from \eqref{eq:SEP_GENERAL}, the SEP depends on the probability of incorrect symbol detection and, similar to the discrete-input discrete-output MI expression given by \eqref{eq::I_DD}, is governed by the distribution of constellation points in 3D space. A well-structured 3D constellation lattice, where symbols remain sufficiently distinguishable in the presence of noise, reduces the probability of symbol misclassification. By analyzing the dependence of the SEP on the minimum Euclidean distance, minimizing the SEP can also be achieved by maximizing the minimum distance between constellation points. This reinforces the conclusion that optimizing the minimum distance in the 3D grid simultaneously improves both MI and SEP, making it a unified optimization criterion for cross-band modulation design.

Having established the role of minimum Euclidean distance maximization in improving both MI and SEP, we now develop a DNN-based geometric shape optimization framework aimed at enhancing the performance of the proposed cross-band system. The objective is to determine the optimal constellation structure that maximizes the minimum Euclidean distance of the 3D constellation while ensuring compliance with energy constraints. By directly optimizing the constellation geometry, this approach effectively shapes the cross-band signal structure to achieve superior performance in both MI and SEP. In this direction, we propose a DNN architecture to optimize the 3D constellation lattice by learning an improved mapping of RF symbols into the optical domain. The input to the network consists of $M\times2$ nodes, where $M$ represents the constellation order, and each input symbol is characterized by its two RF coordinates, $x_I$ and $x_Q$ components. This input structure ensures that the network directly processes the $M$-QAM symbols, maintaining compatibility with practical RF modulation schemes. The network is composed of three fully connected hidden layers, each containing multiple neurons and using the rectified linear unit (ReLU) activation function to introduce non-linearity and enhance learning capability. The ReLU activation ensures that the network effectively captures complex relationships between the RF input space and the optimal optical intensity mapping while maintaining efficient gradient propagation during training. The output layer consists of $M$ nodes corresponding to the intensity levels of the optical link. To ensure that all output values remain strictly positive, the outputs are passed through a Softplus activation function, which enforces positivity while allowing for smooth optimization. However, learning the optimal mapping $F(x_I, x_Q)$ requires not only an effective network architecture, but also a well-designed loss function that guides the optimization process. To this end, we formulate a custom loss function that explicitly incorporates the maximization of the minimum Euclidean distance between constellation points while ensuring compliance with energy constraints. 

The first component of the loss function is designed to maximize the minimum Euclidean distance $\dmin$ between constellation points. A direct approach to maximizing $\dmin$ would involve solving a max-min optimization problem, which is inherently non-differentiable and difficult to handle in gradient-based training. To overcome this, we employ a smooth, differentiable surrogate by leveraging the sum of exponentials of the pairwise distances between constellation points. Specifically, we define the distance-based loss function as
\begin{equation} \label{eq:loss_function_max_dmin}
    \begin{aligned}
    \mathcal{L}_{d} = \sum_{j=1}^{M}\sum_{\substack{i=1 \\ i \neq j}}^{M}e^{-\kappa d_{i}\left(x_{I,j}, x_{Q,j}, z_{O,j}\right)},
    \end{aligned}
\end{equation}
where $d_{i}\left(x_{I,j}, x_{Q,j}, z_{O,j}\right)$ is the pairwise weighted Euclidean distance between the $i$-th and the $j$-th symbol of the 3D constellation, calculated by \eqref{dmin}. Here, $z_{O,j}$ represents the intensity value output by the DNN for the $j$-th symbol, and $\kappa$ is a hyperparameter that controls the sensitivity of the loss function to distance variations.
The intuition behind this formulation is that when the minimum distance $\dmin$ is large, at least one term in the summation will dominate and drive the total sum to lower values. Conversely, when $\dmin$ is small, the corresponding exponential term grows significantly, increasing the overall loss function. By minimizing this loss function during training, we inherently push the constellation points further apart, ensuring that the network learns a constellation structure that maximizes the minimum Euclidean distance. Thus, minimizing $\mathcal{L}_d$ is equivalent to solving the original max-min problem in a differentiable way, making it suitable for gradient-based learning.

In addition to maximizing the minimum weighted Euclidean distance given by \eqref{dmin}, the learned constellation must satisfy energy constraints to ensure practical feasibility in the optical transmission system. To enforce this constraint, we introduce a second loss term that penalizes deviations from the desired average power level. This energy constraint is formulated as
\begin{equation}\label{eq:loss_function_energy}
    \begin{aligned}
        \mathcal{L}_e = \left(\frac{1}{M}\sum_{j=1}^{M}z_{O,j} - 1\right)^2.
    \end{aligned}
\end{equation}
This term ensures that the average optical power of the learned constellation is unity. Any deviation from this target results in an increase in the loss function, encouraging the network to adjust the constellation accordingly.
The final loss function combines \eqref{eq:loss_function_max_dmin} and \eqref{eq:loss_function_energy}, balancing the maximization of the minimum weighted Euclidean distance with the enforcement of the power constraint and is expressed as
\begin{equation} \label{eq:loss_function}
    \begin{aligned}
        \mathcal{L} =\mathcal{L}_d + \lambda\mathcal{L}_e,
    \end{aligned}
\end{equation}
where $\lambda$ acts as a penalty factor that regulates the trade-off between maximizing the minimum distance and maintaining power constraints. A higher $\lambda$ places greater emphasis on meeting the energy constraint, while a lower value prioritizes distance maximization. By minimizing this combined loss function, the network simultaneously pushes the constellation points apart while ensuring that the average power remains within the required limits, leading to an optimized 3D lattice that enhances both MI and SEP.

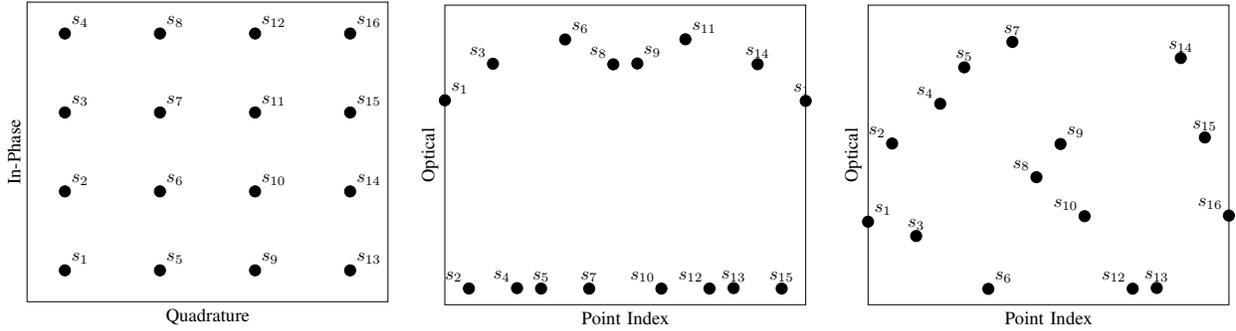
\begin{figure*}
\centering
\begin{minipage}[t]{0.3\textwidth}
\centering
\begin{tikzpicture}[scale=0.7]
\begin{axis}[
        xtick=\empty,
        ytick=\empty,
        xlabel={$\text{Quadrature}$},
        ylabel={$\text{In-Phase}$},
        xlabel style={below},
        ylabel style={above},
        xmin=-1.2,
        xmax=1.2,
        ymin=-1.2,
        ymax=1.2
    ]
    \addplot [only marks, mark size = 3,black] table {figures/3D_constellations/16_QAM_rf_input.txt};
    \node at (-0.948683, -0.948683) [anchor=south west] {$s_1$};
    \node at (-0.948683, -0.316228) [anchor=south west] {$s_2$};
    \node at (-0.948683, 0.316228) [anchor=south west] {$s_3$};
    \node at (-0.948683, 0.9486830) [anchor=south west] {$s_4$};
    \node at (-0.316228, -0.948683) [anchor=south west] {$s_5$};
    \node at (-0.316228, -0.316228) [anchor=south west] {$s_6$};
    \node at (-0.316228, 0.316228) [anchor=south west] {$s_7$};
    \node at (-0.316228, 0.948683) [anchor=south west] {$s_8$};
    \node at (0.316228,	-0.948683) [anchor=south west] {$s_9$};
    \node at (0.316228, -0.316228) [anchor=south west] {$s_{10}$};
    \node at (0.316228, 0.316228) [anchor=south west] {$s_{11}$};
    \node at (0.316228, 0.948683) [anchor=south west] {$s_{12}$};
    \node at (0.948683, -0.948683) [anchor=south west] {$s_{13}$};
    \node at (0.948683, -0.316228) [anchor=south west] {$s_{14}$};
    \node at (0.948683, 0.316228) [anchor=south west] {$s_{15}$};
    \node at (0.948683, 0.948683) [anchor=south west] {$s_{16}$};

\end{axis}
\end{tikzpicture}
\label{fig:QAM-iq}
\end{minipage}
\begin{minipage}[t]{0.3\textwidth}
\centering
\begin{tikzpicture}[scale=0.7]
\begin{axis}[
        xtick=\empty,
        ytick=\empty,
        xlabel={$\text{Point Index}$ },
        ylabel={$\text{Optical}$},
        xlabel style={below},
        ylabel style={above},
        xmin=1,
        xmax=16,
        ymin=-0.1,
        ymax=1.8
    ]
    \addplot [only marks, mark size = 3,black] table {figures/3D_constellations/optimized_z_snr_optical_0dB_snr_rf_10dB.txt};
    \node at (1, 1.19651) [anchor=south west] {$s_1$};
    \node at (2, 0.0029704) [anchor=south east] {$s_2$};
    \node at (3, 1.42842) [anchor=south east] {$s_3$};
    \node at (4, 0.00521091) [anchor=south east] {$s_4$};
    \node at (5, 0.0026) [anchor=south] {$s_5$};
    \node at (6, 1.58236) [anchor=south west] {$s_6$};
    \node at (7, 0.00209996) [anchor=south] {$s_7$};
    \node at (8, 1.42474) [anchor=south east] {$s_8$};
    \node at (9, 1.42997) [anchor=south west] {$s_9$};
    \node at (10, 0.00221412) [anchor=south east] {$s_{10}$};
    \node at (11, 1.58357) [anchor=south west] {$s_{11}$};
    \node at (12, 0.00266886) [anchor=south east] {$s_{12}$};
    \node at (13, 0.00550037) [anchor=south] {$s_{13}$};
    \node at (14, 1.42543) [anchor=south] {$s_{14}$};
    \node at (15, 0.00267423) [anchor=south] {$s_{15}$};
    \node at (16, 1.19283) [anchor=south] {$s_{16}$};
\end{axis}
\end{tikzpicture}
\label{fig:QAM-polar}
\end{minipage}
\begin{minipage}[t]{0.3\textwidth}
\centering
\begin{tikzpicture}[scale=0.7]
\begin{axis}[
        xtick=\empty,
        ytick=\empty,
        xlabel={$\text{Point Index}$ },
        ylabel={$\text{Optical}$},
        xlabel style={below},
        ylabel style={above},
        xmin=1,
        xmax=16,
        ymin=-0.1,
        ymax=1.8
    ]
    \addplot [only marks, mark size = 3,black] table {figures/3D_constellations/optimized_z_snr_optical_20dB_snr_rf_12dB.txt};
    \node at (1, 0.426533) [anchor=south west] {$s_1$};
    \node at (2, 0.922443) [anchor=south east] {$s_2$};
    \node at (3, 0.335901) [anchor=south] {$s_3$};
    \node at (4, 1.17457) [anchor=south east] {$s_4$};
    \node at (5, 1.40566) [anchor=south] {$s_5$};
    \node at (6, 0.000493295) [anchor=south west] {$s_6$};
    \node at (7, 1.56695) [anchor=south] {$s_7$};
    \node at (8, 0.708897) [anchor=south east] {$s_8$};
    \node at (9, 0.91945) [anchor=south west] {$s_9$};
    \node at (10, 0.461547) [anchor=south east] {$s_{10}$};
    \node at (11, 2.05255) [anchor=south west] {$s_{11}$};
    \node at (12, 0.00189263) [anchor=south east] {$s_{12}$};
    \node at (13, 0.00614749) [anchor=south] {$s_{13}$};
    \node at (14, 1.46465) [anchor=south] {$s_{14}$};
    \node at (15, 0.961093) [anchor=south] {$s_{15}$};
    \node at (16, 0.465358) [anchor=south east] {$s_{16}$};
    \end{axis}
    \end{tikzpicture}   
\end{minipage}
\caption{a) RF input 16-QAM, b) Outputted optical values for $\gamma_1^2 = 10$ dB and $\gamma_2^2 = 0$ dB, c) Outputted optical values for $\gamma_1^2 = 12$ dB and $\gamma_2^2 = 20$ dB.}
\label{fig:optimized_scatter}
\end{figure*}

The impact of the proposed DNN-based optimization can be further understood by examining the learned 3D constellation structures under different RF and optical SNR conditions, as shown in Fig. \ref{fig:optimized_scatter}. First, in Fig. \ref{fig:optimized_scatter}a, we present the input 16-QAM RF constellation, which serves as the input to the neural network, ensuring compatibility with practical $M$-QAM-based RF transmission schemes. Furthermore, Figs. \ref{fig:optimized_scatter}b and \ref{fig:optimized_scatter}c depict the corresponding optimized optical intensity values after DNN-based optimization for low and high optical SNR scenarios, respectively. When the optical SNR is low, the network learns to map the optical intensity values into two well-separated clusters. This clustering structure improves symbol discriminability in the optical domain, ensuring that the received symbols remain resolvable with high reliability despite the low optical SNR. In addition, in this low-SNR scenario, adjacent RF symbols are intentionally separated in the optical domain to prevent adjacent RF symbols from overlapping in the optical intensity dimension, effectively minimizing cross-domain error propagation. As the optical SNR increases, the optimized constellation structure adapts to better utilize the additional channel capacity. Specifically, in Fig. \ref{fig:optimized_scatter}c, we observe that the DNN now assigns a wider range of unique optical intensity levels to the RF symbols, taking advantage of the improved optical SNR to encode more information over the optical link. In addition, unlike the low-SNR case, adjacent RF symbols no longer remain separated in the optical axis, as the system can now afford to maintain finer granularity in symbol placement without compromising reliability. These results highlight the adaptive nature of the learned mapping $F(x_I, x_Q)$ and demonstrate that the DNN dynamically optimizes the 3D constellation structure based on the underlying SNR conditions to jointly improve both MI and SEP performance.

\section{Numerical Results}
In this section, we present numerical results to evaluate the performance of the proposed linear and DNN-Gen cross-band modulation schemes and to validate our analytical derivations. The results are obtained through Monte Carlo simulations with $10^{7}$ symbols, considering a constellation order of $M=16$. To evaluate the effectiveness of the proposed approaches, we compare them with CB-PAM from \cite{sotiris-panos} and magnitude-based cross-band modulation (MCBM) inspired by \cite{Popovski}. The evaluation focuses on both MI and SEP performance, under various RF and optical SNR conditions. In particular, we examine continuous-input schemes, where the trade-offs between Gaussian and exponential RF input distributions affect MI performance, and discrete-input schemes, where the impact of optimized linear mapping and learned DNN-Gen 3D constellations is assessed. The results offer valuable insights into the performance improvements, complexity considerations, and adaptability of the proposed methods, emphasizing their advantages over conventional cross-band modulation schemes.

\begin{figure}
\centering
\begin{minipage}{.5\textwidth}
\centering
\begin{tikzpicture}
 \begin{axis}[
            width=\linewidth,
            xlabel = $\gamma_1^2$ (dB),
            ylabel = Mutual Information,
            xmin = -20,
            xmax = 20,
            ymin = 0,
            ymax = 6,
            xtick = {-20,-10,...,30},
            ytick = {0,2,...,10},
            grid = major,
            legend cell align = {left},
            legend pos = north west,
            legend style={font=\tiny}
            ]
            \addplot[
            color= black,
            line width = 1pt,
		style = dashdotted,
            mark options = solid,
            mark=o,
            mark repeat = 2,
            mark size = 3,
            ]
            table {figures/MI/discrete/linear_optical_snr_10dB.txt};
            \addlegendentry{Linear Mapping}
            
            \addplot[
            color= black,
            line width = 1pt,
		style = dashdotted,
            mark options = solid,
            mark=square,
            mark repeat = 2,
            mark size = 3,
            ]
            table {figures/MI/discrete/magnitude_optical_snr_10dB.txt};
            \addlegendentry{MCBM}

            \addplot[
            color= black,
            line width = 1pt,
		style = dashdotted,
            mark=+,
            mark options = solid,
            mark repeat = 2,
            mark size = 3,
            ]
            table {figures/MI/discrete/optimal_3D_optical_snr_10dB.txt};
            \addlegendentry{DNN-Gen}

            \addplot[
            color= black,
            line width = 1pt,
		style = dashdotted,
            mark=asterisk,
            mark options = solid,
            mark repeat = 2,
            mark size = 3,
            ]
            table {figures/MI/discrete/pam_1D_optical_snr_10dB.txt};
            \addlegendentry{CB-PAM}
            
            \addplot[
            color= black,
            line width = 1pt,
		style = solid,
            mark=,
            mark options = solid,
            mark repeat = 2,
            mark size = 3,
            ]
            table {figures/MI/continuous/linear_theoretical_optical_snr_10dB.txt};
            \addlegendentry{LGCB}

            \addplot[
            color= black,
            line width = 1pt,
		style = solid,
            mark=triangle,
            mark options = solid,
            mark repeat = 2,
            mark size = 3,
            ]
            table {figures/MI/continuous/gamma_linear_theoretical_optical_snr_10dB.txt};
            \addlegendentry{LXCB}

\end{axis}
\end{tikzpicture}
\subcaption{$\gamma_2^2 = 10$ dB}
\end{minipage}
\begin{minipage}{.5\textwidth}
\centering
\begin{tikzpicture}
 \begin{axis}[
            width=\linewidth,
            xlabel = $\gamma_1^2$ (dB),
            ylabel = Mutual Information,
            xmin = -20,
            xmax = 20,
            ymin = 0,
            ymax = 6,
            xtick = {-20,-10,...,30},
            ytick = {0,2,...,10},
            grid = major,
            legend cell align = {left},
            legend pos = south east,
            legend style={font=\tiny}
            ]
        \addplot[
            color= black,
            line width = 1pt,
		style = dashdotted,
            mark=o,
            mark options = solid,
            mark repeat = 2,
            mark size = 3,
            ]
            table {figures/MI/discrete/linear_optical_snr_20dB.txt};
            \addlegendentry{Linear Mapping}
            
            \addplot[
            color= black,
            line width = 1pt,
		style = dashdotted,
            mark=square,
            mark options = solid,
            mark repeat = 2,
            mark size = 3,
            ]
            table {figures/MI/discrete/magnitude_optical_snr_20dB.txt};
            \addlegendentry{MCBM}

            \addplot[
            color= black,
            line width = 1pt,
		style = dashdotted,
            mark=+,
            mark options = solid,
            mark repeat = 2,
            mark size = 3,
            ]
            table {figures/MI/discrete/optimal_3D_optical_snr_20dB.txt};
            \addlegendentry{DNN-Gen}

            \addplot[
            color= black,
            line width = 1pt,
		style = dashdotted,
            mark=asterisk,
            mark options = solid,
            mark repeat = 2,
            mark size = 3,
            ]
            table {figures/MI/discrete/pam_1D_optical_snr_20dB.txt};
            \addlegendentry{CB-PAM}
            
            \addplot[
            color= black,
            line width = 1pt,
		style = solid,
            mark=,
            mark options = solid,
            mark repeat = 2,
            mark size = 3,
            ]
            table {figures/MI/continuous/linear_theoretical_optical_snr_20dB.txt};
            \addlegendentry{LGCB}

            \addplot[
            color= black,
            line width = 1pt,
		style = solid,
            mark=triangle,
            mark options = solid,
            mark repeat = 2,
            mark size = 3,
            ]
            table {figures/MI/continuous/gamma_linear_theoretical_optical_snr_20dB.txt};
            \addlegendentry{LXCB}

\end{axis}
\end{tikzpicture}
\subcaption{$\gamma_2^2 = 20$ dB}
\end{minipage}
\caption{MI versus received SNR of RF subsystem for various schemes.}
\label{fig:MI}
\end{figure}
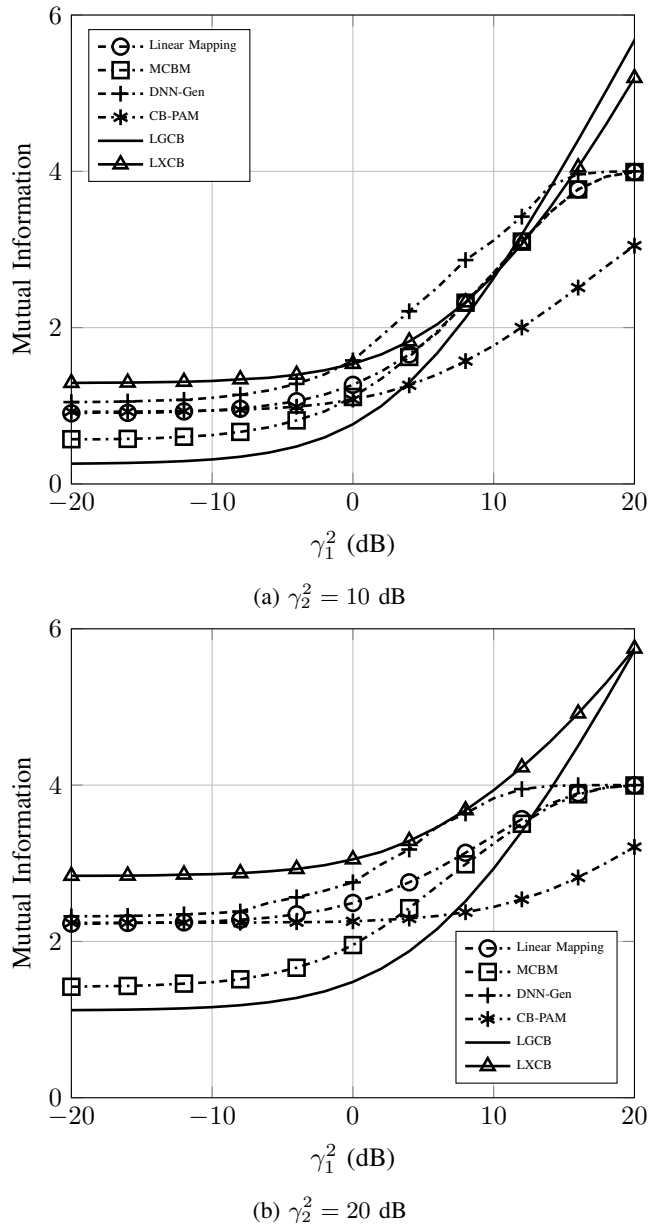

Figs. \ref{fig:MI}a and \ref{fig:MI}b illustrate the MI performance of the benchmark schemes LGCB and LXCB, along with the proposed linear and DNN-generated cross-band modulation schemes, in comparison to conventional cross-band approaches, namely CB-PAM and MCBM. The MI is plotted as a function of the RF SNR under two different optical SNR conditions, with Fig. 3a corresponding to $\gamma_1^2=10$ and Fig. \ref{fig:MI}b corresponding to $\gamma_2^2=20$. Additionally, the performance of the continuous-input schemes, LGCB and LXCB, is presented as a reference to highlight the fundamental trade-offs between Gaussian and exponential input distributions in the proposed cross-band setting. Considering the continuous input schemes, LXCB exhibits superior MI performance at low RF SNRs, which is attributed to its exponential input distribution, which allows better utilization of the optical channel. This advantage is even more pronounced in Fig. \ref{fig:MI}b, where a higher optical SNR further enhances LXCB's ability to efficiently utilize the optical link, widening the MI gap between LXCB and LGCB. However, as the RF SNR increases, LGCB gradually outperforms LXCB. This occurs because Gaussian inputs maximize the MI in the RF link, allowing LGCB to utilize the increased RF SNR more effectively than LXCB, which operates with chi-squared distributed RF inputs due to the transformation of the gamma distributed components.  

Moreover, turning to the discrete-input schemes, both the proposed linear cross-band mapping and the proposed DNN-Gen approach consistently outperform CB-PAM and MCBM, verifying the effectiveness of the proposed approaches. The linear mapping scheme, which utilizes optimized linear coefficients to map RF symbols to optical intensity, provides a simple yet efficient solution that is straightforward to optimize via \eqref{eq:maxmin_with_theta}, while also maintaining an $\mathcal{O}(1)$ detection complexity. The DNN-Gen scheme, on the other hand, further enhances MI by learning an optimized nonlinear mapping function, improving symbol distinguishability across both the RF and optical domains, and adapting the 3D lattice structure to different SNR conditions. An interesting observation from Fig. \ref{fig:MI}a is that the DNN-generated cross-band scheme outperforms the continuous LGCB at low and medium RF SNRs, despite operating with a discrete input. This demonstrates the effectiveness of learning-based optimization in structuring the 3D constellation for better MI performance. However, as the optical SNR increases in Fig. \ref{fig:MI}b, the LXCB begins to dominate as it leverages its continuous exponential mapping to optimally utilize the high optical SNR, further validating the advantage of exponential input distributions in such conditions. Furthermore, a key limitation of MCBM is evident in Fig. \ref{fig:MI}b, where its MI saturates at high optical SNRs due to its magnitude-based mapping, which assigns multiple RF symbols to the same optical intensity level. This prevents the MCBM from fully utilizing the capacity of the optical channel, resulting in performance saturation. These results highlight the adaptability and robustness of the proposed cross-band modulation schemes and demonstrate their ability to outperform conventional approaches by utilizing either optimized linear or DNN-Gen mappings, which significantly enhance information in cross-band systems.

\begin{figure}
    \centering
    \begin{tikzpicture}
        \begin{semilogyaxis}[
            width=\linewidth,
            xlabel = $\gamma_1^2$ (dB),
            ylabel = Symbol Error Probability,
            xmin = 0,
            xmax = 25,
            ymin = 1e-5,
            ymax = 1,
            xtick = {0,5,...,30},
            grid = major,
            legend cell align = {left},
            legend pos = south west,
            legend style={font=\small}
            ]
        
            \addplot[
            color= black,
            only marks,
            mark=*,
            mark repeat = 1,
            mark size = 2,
            ]
            table {figures/SEP/proposed_optical_snr_10dB.txt};
            \addlegendentry{$\gamma_2 = 10$ dB}
            
            \addplot[
            color= black,
            only marks,
            mark=triangle*,
            mark repeat = 1,
            mark size = 3,
            ]
            table {figures/SEP/proposed_optical_snr_20dB.txt};
            \addlegendentry{$\gamma_2 = 20$ dB}

            \addplot[
            color= black,
            only marks,
            mark=diamond*,
            mark repeat = 1,
            mark size = 3,
            ]
            table {figures/SEP/proposed_optical_snr_25dB.txt};
            \addlegendentry{$\gamma_2 = 25$ dB}
            
            \addplot[
            color= black,
            only marks,
            mark=square*,
            mark repeat = 1,
            mark size = 2,
            ]
            table {figures/SEP/proposed_optical_snr_30dB.txt};
            \addlegendentry{$\gamma_2 = 30$ dB}
            \addplot[
            color= black,
            no marks,
            line width = 1pt,
		style = solid,
            mark=square*,
            mark repeat = 1,
            mark size = 2,
            ]
            table {figures/SEP/proposed_optical_snr_10dB_theory.txt};
            \addlegendentry{SEP Approx. \eqref{SEP_approx}}
            \addplot[
            color= black,
            no marks,
            line width = 1pt,
		style = solid,
            mark=square*,
            mark repeat = 1,
            mark size = 2,
            ]
            table {figures/SEP/proposed_optical_snr_20dB_theory.txt};
            \addplot[
            color= black,
            no marks,
            line width = 1pt,
		style = solid,
            mark=square*,
            mark repeat = 1,
            mark size = 2,
            ]
            table {figures/SEP/proposed_optical_snr_25dB_theory.txt};
            \addplot[
            color= black,
            no marks,
            line width = 1pt,
		style = solid,
            mark=square*,
            mark repeat = 1,
            mark size = 2,
            ]
            table {figures/SEP/proposed_optical_snr_30dB_theory.txt};
        \end{semilogyaxis}
    \end{tikzpicture}
    \vspace{-2mm}
\caption{SEP approximation of the proposed  cross-band modulation with linear mapping.}
\label{fig:SEP_approx}
\end{figure}
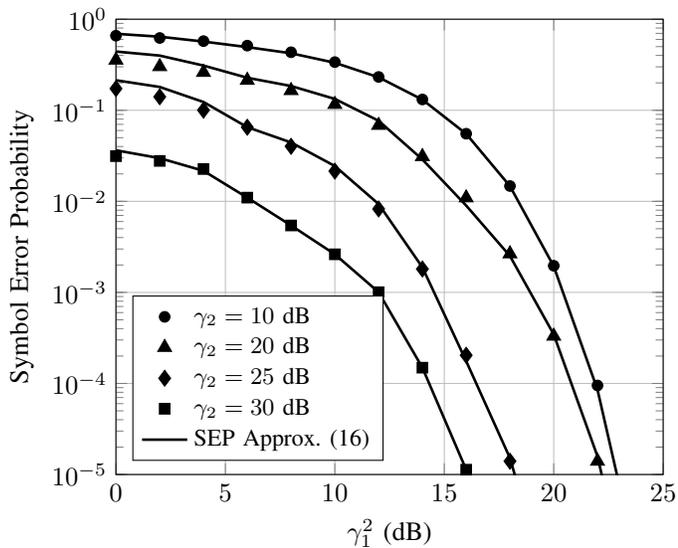

In Fig. \ref{fig:SEP_approx}, we present the SEP approximation derived in \eqref{SEP_approx} for different optical SNR values, specifically $\gamma_2^2\in\{10,20, 25, 30\}$ dB, thus validating its accuracy over all investigated SNR conditions of the cross-band system. For each SNR level, the optimal pair $(a_1,a_2)$ given by \eqref{eq:maxmin_with_theta} is used to generate the corresponding planar 3D constellation. Furthermore, the figure shows the significant performance improvements achieved by the proposed cross-band modulation scheme as the optical SNR increases. These improvements highlight the ability of the proposed cross-band modulation to effectively utilize the diversity gains provided by the optical link, thereby significantly improving the overall performance and capacity of the system.

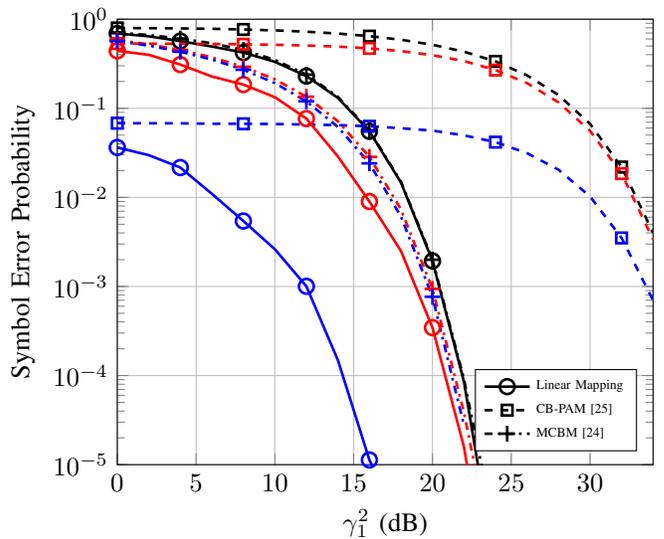
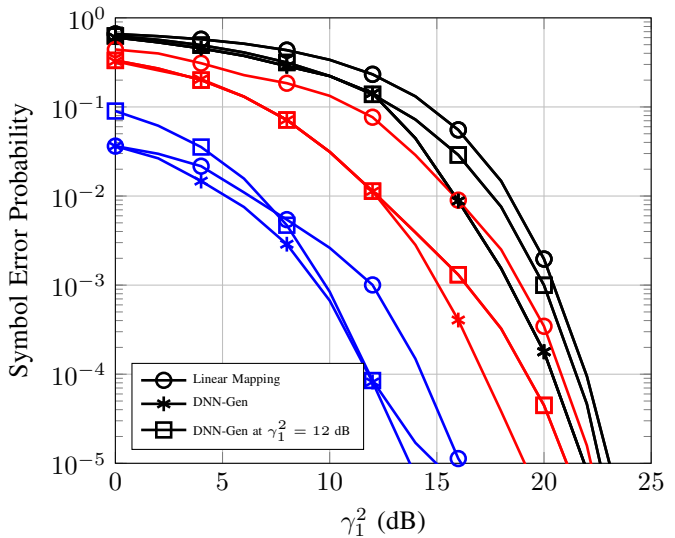
\begin{figure}
\centering
\begin{minipage}{.48\textwidth}
\centering
    \begin{tikzpicture}
        \begin{semilogyaxis}[
            width=\linewidth,
            xlabel = $\gamma_1^2$ (dB),
            ylabel = Symbol Error Probability,
            xmin = 0,
            xmax = 34,
            ymin = 1e-5,
            ymax = 1,
            xtick = {0,5,...,35},
            grid = major,
            legend cell align = {left},
            legend pos = south east,
            legend style={font=\tiny}
            ]

            \addplot[
            color= black,
            line width = 1pt,
		style = solid,
            mark=o,
            mark options = solid,
            mark repeat = 2,
            mark size = 2.8,
            ]
            table {figures/SEP/proposed_optical_snr_10dB_theory.txt};
            \addlegendentry{Linear Mapping}

            \addplot[
            color= black,
            line width = 1pt,
		style = dashed,
            mark=square,
            mark options = solid,
            mark repeat = 4,
            mark size = 2,
            ]
            table {figures/SEP/pam_optical_snr_10dB.txt};
            \addlegendentry{CB-PAM \cite{sotiris-panos}}

            \addplot[
            color= black,
            line width = 1pt,
		style = dashdotted,
            mark=+,
            mark options = solid,
            mark repeat = 2,
            mark size = 3,
            ]
            table {figures/SEP/popovski_optical_snr_10dB.txt};
            \addlegendentry{MCBM \cite{Popovski}}

            \addplot[
            color= red,
            line width = 1pt,
		style = dashed,
            mark=square,
            mark options = solid,
            mark repeat = 4,
            mark size = 2,
            ]
            table {figures/SEP/pam_optical_snr_20dB.txt};
            \addplot[
            color= blue,
            line width = 1pt,
		style = dashed,
            mark=square,
            mark options = solid,
            mark repeat = 4,
            mark size = 2,
            ]
            table {figures/SEP/pam_optical_snr_30dB.txt};

            \addplot[
            color= red,
            line width = 1pt,
		style = solid,
            mark=o,
            mark options = solid,
            mark repeat = 2,
            mark size = 2.8,
            ]
            table {figures/SEP/proposed_optical_snr_20dB_theory.txt};
            \addplot[
            color= blue,
            line width = 1pt,
		style = solid,
            mark=o,
            mark options = solid,
            mark repeat = 2,
            mark size = 2.8,
            ]
            table {figures/SEP/proposed_optical_snr_30dB_theory.txt};

            \addplot[
            color= red,
            line width = 1pt,
		style = dashdotted,
            mark=+,
            mark options = solid,
            mark repeat = 2,
            mark size = 3,
            ]
            table {figures/SEP/popovski_optical_snr_20dB.txt};
            \addplot[
            color= blue,
            line width = 1pt,
		style = dashdotted,
            mark=+,
            mark options = solid,
            mark repeat = 2,
            mark size = 3,
            ]
            table {figures/SEP/popovski_optical_snr_30dB.txt};
        \end{semilogyaxis}
    \end{tikzpicture}
\subcaption{Linear Mapping}
\end{minipage}
\begin{minipage}{.48\textwidth}
\centering
    \begin{tikzpicture}
        \begin{semilogyaxis}[
            width=\linewidth,
            xlabel = $\gamma_1^2$ (dB),
            ylabel = Symbol Error Probability,
            xmin = 0,
            xmax = 25,
            ymin = 1e-5,
            ymax = 1,
            xtick = {0,5,...,25},
            grid = major,
            legend cell align = {left},
            legend pos = south west,
            legend style={font=\tiny}
            ]

            \addplot[
            color= black,
            line width = 1pt,
		style = solid,
            mark=o,
            mark options = solid,
            mark repeat = 2,
            mark size = 2.8,
            ]
            table {figures/SEP/proposed_optical_snr_10dB.txt};
            \addlegendentry{Linear Mapping}

            \addplot[
            color= black,
            line width = 1pt,
		style = solid,
            mark=asterisk,
            mark options = solid,
            mark repeat = 2,
            mark size = 2.8,
            ]
            table {figures/SEP/optimal_optical_snr_10dB.txt};
            \addlegendentry{DNN-Gen}
            \addplot[
            color= black,
            line width = 1pt,
		style = solid,
            mark=square,
            mark options = solid,
            mark repeat = 2,
            mark size = 2.8,
            ]
            table {figures/SEP/optimal_optical_snr_10dB_rf_snr_10dB.txt};
            \addlegendentry{DNN-Gen at $\gamma_1^2 = 12$ dB}

            \addplot[
            color= black,
            line width = 1pt,
		style = solid,
            mark=o,
            mark options = solid,
            mark repeat = 2,
            mark size = 2.8,
            ]
            table {figures/SEP/proposed_optical_snr_10dB.txt};

            \addplot[
            color= black,
            line width = 1pt,
		style = solid,
            mark=asterisk,
            mark options = solid,
            mark repeat = 2,
            mark size = 2.8,
            ]
            table {figures/SEP/optimal_optical_snr_10dB.txt};

            \addplot[
            color= black,
            line width = 1pt,
		style = solid,
            mark=square,
            mark options = solid,
            mark repeat = 2,
            mark size = 2.8,
            ]
            table {figures/SEP/optimal_optical_snr_10dB_rf_snr_10dB.txt};
            
            
            \addplot[
            color= red,
            line width = 1pt,
		style = solid,
            mark=square,
            mark options = solid,
            mark repeat = 2,
            mark size = 2.8,
            ]
            table {figures/SEP/optimal_optical_snr_20dB_rf_snr_12dB.txt};

            \addplot[
            color= red,
            line width = 1pt,
		style = solid,
            mark=square,
            mark options = solid,
            mark repeat = 2,
            mark size = 2.8,
            ]
            table {figures/SEP/optimal_optical_snr_20dB_rf_snr_12dB.txt};

            \addplot[
            color= red,
            line width = 1pt,
		style = solid,
            mark=o,
            mark options = solid,
            mark repeat = 2,
            mark size = 2.8,
            ]
            table {figures/SEP/proposed_optical_snr_20dB_theory.txt};

            \addplot[
            color= black,
            line width = 1pt,
		style = solid,
            mark=asterisk,
            mark options = solid,
            mark repeat = 2,
            mark size = 2.8,
            ]
            table {figures/SEP/optimal_optical_snr_10dB.txt};

            \addplot[
            color= red,
            line width = 1pt,
		style = solid,
            mark=asterisk,
            mark options = solid,
            mark repeat = 2,
            mark size = 2.8,
            ]
            table {figures/SEP/optimal_optical_snr_20dB.txt};

            \addplot[
            color= blue,
            line width = 1pt,
		style = solid,
            mark=asterisk,
            mark options = solid,
            mark repeat = 2,
            mark size = 2.8,
            ]
            table {figures/SEP/optimal_optical_snr_30dB.txt};

            \addplot[
            color= blue,
            line width = 1pt,
		style = solid,
            mark=o,
            mark options = solid,
            mark repeat = 2,
            mark size = 2.8,
            ]
            table {figures/SEP/proposed_optical_snr_30dB_theory.txt};

            \addplot[
            color= blue,
            line width = 1pt,
		style = solid,
            mark=square,
            mark options = solid,
            mark repeat = 2,
            mark size = 2.8,
            ]
            table {figures/SEP/optimal_optical_snr_30dB_rf_snr_12dB.txt};     
        \end{semilogyaxis}
    \end{tikzpicture}
\subcaption{DNN-Gen}
\end{minipage}
\caption{SEP versus received SNR of RF subsystem for $\gamma_2^2 = 10$ dB (black), $\gamma_2^2 = 20$ dB (red), and $\gamma_2^2 = 30$ dB (blue).}
\label{fig:compare}
\end{figure}

In Fig. \ref{fig:compare}a, we compare the SEP performance of the proposed linear mapping-based cross-band modulation scheme, optimized at each SNR, against CB-PAM and MCBM over optical SNR values of $\gamma_2^2\in\{10,20,30\}$ dB. The proposed scheme consistently outperforms CB-PAM over all tested SNR conditions and shows significant gains over MCBM, especially at higher optical SNR levels. While the proposed system and MCBM demonstrate comparable performance at low optical SNR, the proposed system achieves approximately $1$ dB SNR gains for a target SEP of $10^{-3}$ at moderate SNR levels. Furthermore, as the optical SNR increases, the performance gap between the proposed modulation and MCBM widens, highlighting the effectiveness of the optimized linear mapping at higher SNR regimes.
A key observation in Fig. \ref{fig:compare}a is that as the optical SNR increases, the SEP decreases even in the low RF SNR regime. This effect occurs because the M-PAM structure created by the optimized linear mapping coefficients $(a_1, a_2)$ in the optical domain increases the reliability of information transmission. As a result, the blue curve ($\gamma_2^2 = 30$ dB) starts below $10^{-1}$ even at $\gamma_1^2 = 0$ dB, demonstrating that at high optical SNRs, the system can maintain relatively low error rates even under weak RF conditions. This advantage is further enhanced by the inherent limitations of MCBM. Due to its magnitude-based mapping, MCBM assigns multiple RF symbols to the same optical intensity level, limiting its ability to fully exploit the capacity of the optical link. As a result, its performance saturates at $\gamma_1=20$ dB and $\gamma_1=30$, preventing further error rate improvements even as the optical SNR increases. In contrast, the proposed linear mapping cross-band modulation benefits from a simple yet effective one-to-one mapping between RF and optical symbols, allowing it to fully leverage the optical SNR gains without introducing additional complexity. This straightforward mapping ensures that the system can efficiently translate improvements in optical SNR into significant SEP performance gains, especially as SNR conditions improve. By fully utilizing the diversity of the optical link, the proposed linear mapping cross-band modulation achieves significant improvements in overall system SEP performance while maintaining a low-complexity structure that facilitates practical deployment.

Fig. \ref{fig:compare}b shows the SEP performance comparison between the two proposed cross-band schemes: the linear mapping-based scheme and the DNN-Gen scheme, over three different optical SNR conditions $\gamma_2^2\in\{10,20,30\}$ dB. The linear mapping-based scheme, which optimizes the mapping coefficients through a simple procedure, provides an efficient and practical solution with $\mathcal{O}(1)$ detection complexity, making it particularly suitable for resource-constrained implementations. However, the DNN-Gen scheme sacrifices some of this simplicity to generate optimized 3D lattices that aim to maximize MI and minimize the SEP. The effectiveness of this learned mapping is evident in Fig. \ref{fig:compare}b, where DNN-Gen consistently outperforms the linear mapping approach over all tested optical SNR levels, providing performance gains of more than $2$ dB. The figure also examines the performance of DNN-Gen when trained at a specific RF SNR, $\gamma_1^2 = 12$ dB, and evaluated over the entire RF SNR range. Interestingly, the results show that DNN-Gen optimized at a specific RF SNR remains effective over a range of neighboring SNR values, achieving SEP performance comparable to the optimally trained model for each SNR point. This observation suggests a practical advantage for real-world implementations, as it implies that instead of storing a separate 3D lattice for each possible RF SNR condition, a sparser set of pre-trained DNN-Gen lattices could be used, reducing storage requirements while maintaining high performance over a wide SNR range.

\section{Conclusion}
In this work, we presented a novel cross-band modulation framework that improves communication reliability by leveraging RF and optical bands. We proposed two practical modulation strategies, both employing $M$-QAM in the RF subsystem: a linear cross-band mapping scheme and a DNN-Gen 3D constellation, both designed to improve MI and SEP. The linear mapping scheme provides a low-complexity, tractable optimization approach where optimal coefficients minimize the SEP while ensuring $\mathcal{O}(1)$ detection complexity, making it suitable for practical use. In addition, our theoretical MI and SEP analysis provides fundamental insights beyond the linear case, contributing to the broader framework of cross-band modulation. To further optimize performance, we introduced a DNN-generated cross-band modulation that learns optimized 3D lattices, achieving gains of more than 2 dB over the linear approach. Simulation results confirmed that both proposed schemes outperform existing cross-band modulation techniques such as CB-PAM and MCBM, demonstrating their ability to efficiently map RF symbols to the optical domain and fully exploit cross-band diversity. These results establish the proposed framework as a scalable and high-performance solution for future hybrid RF-optical communication systems.

\section*{Acknowledgment}
The authors would like to thank Nikos A. Mitsiou for his valuable discussions and insights that contributed to the development of this work. His feedback and suggestions helped refine key aspects of this study.
\bibliographystyle{IEEEtran} 
\bibliography{IEEEabrv, Bibliography.bib}

\begin{thebibliography}{10}
\providecommand{\url}[1]{#1}
\csname url@samestyle\endcsname
\providecommand{\newblock}{\relax}
\providecommand{\bibinfo}[2]{#2}
\providecommand{\BIBentrySTDinterwordspacing}{\spaceskip=0pt\relax}
\providecommand{\BIBentryALTinterwordstretchfactor}{4}
\providecommand{\BIBentryALTinterwordspacing}{\spaceskip=\fontdimen2\font plus
\BIBentryALTinterwordstretchfactor\fontdimen3\font minus \fontdimen4\font\relax}
\providecommand{\BIBforeignlanguage}[2]{{%
\expandafter\ifx\csname l@#1\endcsname\relax
\typeout{** WARNING: IEEEtran.bst: No hyphenation pattern has been}%
\typeout{** loaded for the language `#1'. Using the pattern for}%
\typeout{** the default language instead.}%
\else
\language=\csname l@#1\endcsname
\fi
#2}}
\providecommand{\BIBdecl}{\relax}
\BIBdecl

\bibitem{6G_karag}
Z.~Zhang, Y.~Xiao, Z.~Ma, M.~Xiao, Z.~Ding, X.~Lei, G.~K. Karagiannidis, and P.~Fan, ``{6G} wireless networks: Vision, requirements, architecture, and key technologies,'' \emph{IEEE Veh. Technol. Mag.}, vol.~14, no.~3, pp. 28--41, 2019.

\bibitem{6G}
B.~Rong, ``{6G}: The next horizon: From connected people and things to connected intelligence,'' \emph{IEEE Wireless Commun.}, vol.~28, no.~5, pp. 8--8, 2021.

\bibitem{rapaport}
T.~Rappaport, \emph{Wireless Communications: Principles and Practice}.\hskip 1em plus 0.5em minus 0.4em\relax USA: Prentice Hall PTR, 2001.

\bibitem{ow-1}
H.~Elgala, R.~Mesleh, and H.~Haas, ``Indoor optical wireless communication: potential and state-of-the-art,'' \emph{IEEE Commun. Mag.}, vol.~49, no.~9, pp. 56--62, 2011.

\bibitem{ow-2}
M.~A. Khalighi and M.~Uysal, ``Survey on free space optical communication: A communication theory perspective,'' \emph{IEEE Commun. Surveys Tuts.}, vol.~16, no.~4, pp. 2231--2258, 2014.

\bibitem{ow-3}
Z.~Ghassemlooy, S.~Arnon, M.~Uysal, Z.~Xu, and J.~Cheng, ``Emerging optical wireless communications-advances and challenges,'' \emph{IEEE J. Sel. Areas Commun.}, vol.~33, no.~9, pp. 1738--1749, 2015.

\bibitem{ow-4}
P.~H. Pathak, X.~Feng, P.~Hu, and P.~Mohapatra, ``Visible light communication, networking, and sensing: A survey, potential and challenges,'' \emph{IEEE Commun. Surveys Tuts.}, vol.~17, no.~4, pp. 2047--2077, 2015.

\bibitem{ow-5}
M.~Z. Chowdhury, M.~T. Hossan, A.~Islam, and Y.~M. Jang, ``A comparative survey of optical wireless technologies: Architectures and applications,'' \emph{IEEE Access}, vol.~6, pp. 9819--9840, 2018.

\bibitem{hybrid-1}
M.~Z. Chowdhury, M.~K. Hasan, M.~Shahjalal, M.~T. Hossan, and Y.~M. Jang, ``Optical wireless hybrid networks: Trends, opportunities, challenges, and research directions,'' \emph{IEEE Commun. Surveys Tuts}, vol.~22, no.~2, pp. 930--966, 2020.

\bibitem{hybrid-2}
S.~R. Teli, C.~Guerra-Yanez, V.~M. Icaza, R.~Perez-Jimenez, Z.~Ghassemlooy, and S.~Zvanovec, ``Hybrid optical wireless communication for versatile {IoT} applications: Data rate improvement and analysis,'' \emph{IEEE Access}, vol.~11, pp. 55\,107--55\,116, 2023.

\bibitem{hybrid-3}
M.~Z. Chowdhury, M.~K. Hasan, M.~Shahjalal, M.~T. Hossan, and Y.~Min~Jang, ``Optical wireless hybrid networks for {5G} and beyond communications,'' in \emph{Proc. International Conference on Information and Communication Technology Convergence (ICTC)}, 2018, pp. 709--712.

\bibitem{Dobre}
L.~Bariah, L.~Mohjazi, S.~Muhaidat, P.~C. Sofotasios, G.~K. Kurt, H.~Yanikomeroglu, and O.~A. Dobre, ``A prospective look: Key enabling technologies, applications and open research topics in {6G} networks,'' \emph{IEEE Access}, vol.~8, pp. 174\,792--174\,820, 2020.

\bibitem{hybrid-4}
S.~A.~H. Mohsan, M.~A. Khan, and H.~Amjad, ``Hybrid {FSO/RF} networks: A review of practical constraints, applications and challenges,'' \emph{Optical Switching and Networking}, vol.~47, p. 100697, 2023.

\bibitem{hybrid-5}
S.~A.~H. Mohsan and H.~Amjad, ``A comprehensive survey on hybrid wireless networks: practical considerations, challenges, applications and research directions,'' \emph{Optical and Quantum Electronics}, vol.~53, no.~9, p. 523, Aug 2021.

\bibitem{hybrid-6}
M.~Z. Chowdhury, M.~T. Hossan, M.~K. Hasan, and Y.~M. Jang, ``Integrated {RF/Optical} wireless networks for improving {QoS} in indoor and transportation applications,'' \emph{Wireless Personal Communications}, vol. 107, no.~3, pp. 1401--1430, Aug 2019.

\bibitem{Haas}
Y.~{Wang} and H.~{Haas}, ``Dynamic load balancing with handover in hybrid li-fi and wi-fi networks,'' \emph{Journal of Lightwave Technology}, vol.~33, no.~22, pp. 4671--4682, Nov. 2015.

\bibitem{Hranilovic}
W.~{Zhang}, S.~{Hranilovic}, and C.~{Shi}, ``Soft-switching hybrid fso/rf links using short-length raptor codes: Design and implementation,'' \emph{IEEE J. Sel. Areas Commun.}, vol.~27, no.~9, pp. 1698--1708, December 2009.

\bibitem{Nestoras-2}
N.~D. Chatzidiamantis, L.~Georgiadis, H.~G. Sandalidis, and G.~K. Karagiannidis, ``Throughput-optimal link-layer design in power constrained hybrid ow/rf systems,'' \emph{IEEE J. Sel. Areas Commun.}, vol.~33, no.~9, pp. 1972--1984, 2015.

\bibitem{Rallis}
K.~G. Rallis, V.~K. Papanikolaou, P.~D. Diamantoulakis, S.~A. Tegos, A.~A. Dowhuszko, M.-A. Khalighi, and G.~K. Karagiannidis, ``Energy efficient cooperative communications in aggregated {VLC/RF} networks with {NOMA},'' \emph{IEEE Trans. Commun.}, vol.~71, no.~9, pp. 5408--5419, 2023.

\bibitem{Nestoras-1}
N.~D. Chatzidiamantis, G.~K. Karagiannidis, E.~E. Kriezis, and M.~Matthaiou, ``Diversity combining in hybrid {RF/FSO} systems with psk modulation,'' in \emph{Proc. IEEE International Conference on Communications (ICC)}, 2011, pp. 1--6.

\bibitem{Optik}
M.~A. Amirabadi and V.~T. Vakili, ``A novel hybrid {FSO / RF} communication system with receive diversity,'' \emph{Optik}, vol. 184, pp. 293 -- 298, 2019.

\bibitem{Hybrid-Alouini}
R.~Samy, H.-C. Yang, T.~Rakia, and M.-S. Alouini, ``Hybrid {SAG-FSO/SH-FSO/RF} transmission for next-generation satellite communication systems,'' \emph{IEEE Trans. Veh. Technol.}, vol.~72, no.~11, pp. 14\,255--14\,267, 2023.

\bibitem{Yue}
Y.~{Xiao}, P.~D. {Diamantoulakis}, Z.~{Fang}, Z.~{Ma}, L.~{Hao}, and G.~K. {Karagiannidis}, ``Hybrid lightwave/{RF} cooperative {NOMA} networks,'' \emph{IEEE Trans. Wireless Commun.}, pp. 1--1, 2019.

\bibitem{Popovski}
W.~{Liu}, X.~{Zhou}, S.~{Durrani}, and P.~{Popovski}, ``A novel receiver design with joint coherent and non-coherent processing,'' \emph{IEEE Trans. Commun.}, vol.~65, no.~8, pp. 3479--3493, Aug 2017.

\bibitem{sotiris-panos}
S.~A. Tegos and P.~D. Diamantoulakis, ``Cross-band {RF}-lightwave communications with integrated passive receiver,'' \emph{IEEE Commun. Lett.}, vol.~27, no.~10, pp. 2583--2587, 2023.

\bibitem{Information-Theory}
T.~M. Cover and J.~A. Thomas, \emph{Elements of Information Theory (Wiley Series in Telecommunications and Signal Processing)}.\hskip 1em plus 0.5em minus 0.4em\relax USA: Wiley-Interscience, 2006.

\bibitem{wigger}
A.~{Lapidoth}, S.~M. {Moser}, and M.~A. {Wigger}, ``On the capacity of free-space optical intensity channels,'' \emph{IEEE Trans. Inf. Theory}, vol.~55, no.~10, pp. 4449--4461, Oct 2009.

\bibitem{proakis}
P.~Massoud~Salehi and J.~Proakis, \emph{Digital Communications}.\hskip 1em plus 0.5em minus 0.4em\relax McGraw-Hill Education, 2007.

\end{thebibliography}

\end{document}